\documentclass[12pt]{iopart}
\usepackage[utf8]{inputenc}
\usepackage[T1]{fontenc}
\usepackage{iopams,setstack,color,dsfont,graphicx,scalerel,wasysym,bm,bbm}
\usepackage[colorlinks=true,linkcolor=blue,citecolor=blue,urlcolor=blue,bookmarks]{hyperref}
\usepackage{doi}
\usepackage{IOPDefs}
\usepackage{CircuitTikz}
\usepackage[compress]{cite}
\eqnobysec
\begin{document}

\title[Dynamics of symmetry-resolved entanglement in Rule 54]
{Non-equilibrium dynamics of symmetry-resolved entanglement and entanglement
asymmetry: Exact asymptotics in Rule 54}

\author{Katja Klobas}

\address{School of Physics and Astronomy, University of Birmingham, Edgbaston, Birmingham, B15 2TT, UK}
\ead{k.klobas@bham.ac.uk}
\vspace{10pt}
\begin{indented}
\item[]November 2024
\end{indented}

\begin{abstract}
Symmetry resolved entanglement and entanglement asymmetry are two measures of quantum correlations sensitive to symmetries of the system. Here we discuss their non-equilibrium dynamics in the Rule 54 cellular automaton, a simple, yet interacting, integrable model. Both quantities can be expressed in terms of the more analytically tractable ``charged moments'', i.e.\ traces of powers of a suitably deformed density matrix, via a replica trick. We express them in terms of a tensor network, which we contract in space using a system of local algebraic relations. This gives the asymptotic form for the charged moments, valid in the regime of large but finite time that is shorter than all the relevant subsystem sizes. In this regime the charge moments decay exponentially with the rate given by the leading solution to a cubic equation.
\end{abstract}

\smallskip
\noindent
\emph{This paper is dedicated to the memory of Marko Medenjak (1990--2022). In the two years that we overlapped as PhD students and shared the office, he made a deep impact on me in my formative years as a researcher. I have very fond memories of long coffee breaks, during which we finished all the leftover seminar biscuits, and heated physics discussions that led to many scientific bets resulting in free pizza for the whole group. Marko has taught me many useful skills, perhaps the most notable being how to most efficiently roll on the chair from the desk to the whiteboard. He was one of the brightest people I have met, and had a long and successful career ahead of him when he tragically passed away. He will be sorely missed.}
\smallskip

\section{Introduction}\label{sec:intro}
The dynamics of entanglement in non-equilibrium quantum many-body systems show
remarkable universality. For instance, after a quantum
quench~\cite{calabrese2006time,calabrese2005evolution} the entanglement entropy
between a subsystem and the rest typically shows linear growth followed by
saturation at a value extensive in the subsystem size. This happens under very
mild conditions, and can be observed in a variety of systems, ranging from
integrable~\cite{calabrese2005evolution,fagotti2008evolution,alba2017entanglement,alba2018entanglement,alba2019entanglement,lagnese2022entanglement}
to
chaotic~\cite{laeuchli2008spreading,kim2013ballistic,liu2014entanglement,asplund2015entanglement,nahum2017quantum,pal2018entangling,bertini2019entanglement,gopalakrishnan2019unitary,piroli2020exact,zhou2020entanglement}.
The reason for this ubiquity of entanglement growth is intimately connected to
thermalisation: a quantum system equilibrates after the entanglement has spread
through the system, and the entanglement entropy has saturated at a value that
coincides with the thermal entropy of the reduced density matrix. The latter is
typically extensive regardless of the number of (local) conservation laws, and
therefore the same general dependence can emerge for both integrable and
chaotic dynamics.

The standard measure of (bipartite) entanglement is the entanglement entropy,
but there are many other measures of quantum
correlations~\cite{amico2008entanglement,horodecki2009quantum}.  Here we
consider two examples that probe the interplay between entanglement and
symmetries (local conserved charges) of the system. The first one is the
\emph{symmetry resolved entanglement
entropy}~\cite{laflorencie2014spin,goldstein2018symmetry,xavier2018equipartition,bonsignori2019symmetry,murciano2020entanglement},
defined as the entanglement entropy of a symmetric state reduced to a given
symmetry sector. The second one is the \emph{entanglement
asymmetry}~\cite{ares2023entanglement}, given by the relative entropy between a
non-symmetric state and its symmetrised counterpart. Recently, these quantities
have been extensively studied, both theoretically~\cite{laflorencie2014spin,goldstein2018symmetry,xavier2018equipartition,bonsignori2019symmetry,murciano2020entanglement,ares2023entanglement,ares2023lack,capizzi2023entanglement,capizzi2023universal,khor2023confinement,bertini2023nonequilibrium,bertini2024dynamics,murciano2024entanglement,caceffo2024entangled,ferro2024nonequilibrium,rylands2024microscopic,chen2024renyi,banerjee2024symmetry}, and
experimentally~\cite{lukin2019probing,azses2020identification,neven2021symmetry,vitale2022symmetry,rath2023entanglement,joshi2024observing}. One of their appealing features for theoretical
investigation is the fact that they can be --- through a replica trick ---
recast in terms of so-called \emph{charged moments} (see
Sec.~\ref{sec:quantities} for precise definition), which are particularly
convenient for analytical manipulations. Despite that, the vast majority of
exact results come from free theories, as the exact non-equilibrium
dynamics of interacting systems is beyond the reach of existing methods.
Refs.~\cite{bertini2023nonequilibrium,bertini2024dynamics} proposed an
effective description for the dynamics of charged moments in interacting
integrable systems. It is based on the idea of exchanging the roles of space
and time~\cite{bertini2022growth}, which maps the non-equilibrium dynamics of
charged moments to a stationary quantity accessible through standard
tools~\cite{caux2016quench,takahashi1999thermodynamics}. Even though this
approach has quantitative predictive power, it is hard to rigorously prove it
in full generality, and one has to test its predictions against exact results.
Here we provide an example where such exact results can be found in closed
analytical form.

Specifically, we consider the reversible cellular automaton Rule
54~\cite{bobenko1993two}, which is a model on a discrete lattice of qubits
with time-evolution defined in discrete time. In 
recent years this model has been recognised as an example of 
interacting integrable model~\cite{friedman2019integrable,gombor2024integrable}, whose
simplicity allows for exact characterization of various non-equilibrium
properties~\cite{prosen2016integrability,prosen2017exact,inoue2018two,klobas2019time,buca2019exact,klobas2020matrix,klobas2020space,gopalakrishnan2018operator,gopalakrishnan2018hydrodynamics,alba2019operator}
(see also a review~\cite{buca2021rule}). The main technical tool we use here is
a generalization of the techniques introduced in
Refs.~\cite{klobas2021exact,klobas2021exactrelaxation,klobas2021entanglement},
which are based on space-time-duality ideas~\cite{banuls2009matrix,muller2012tensor,hastings2015connecting,bertini2018exact,bertini2019exact,lerose2021influence,lerose2021scaling,sonner2021influence,ippoliti2021postselectionfree,bertini2022entanglement,bertini2022growth}.
In essence, one first recasts the charged moments as a tensor network, and then
contracts it in the \emph{space} direction, which can be done exactly due to a
set of local algebraic relations fulfilled by the model. This procedure gives access to the charged moments for short times, that is, times smaller than the sizes of all the subsystems, reproducing the results of Refs.~\cite{bertini2023nonequilibrium,bertini2024dynamics}.

The rest of the paper is organized as follows. In Sec.~\ref{sec:quantities} we present the precise definitions of the quantities of interest. In Sec.~\ref{sec:setup}, we specialise them to the system and quench protocol considered, and express them in terms of a transfer matrix implementing the evolution in space. In Sec.~\ref{sec:FPs} we characterise the fixed points of this transfer matrix, which encode all the information on the short-time regime. The fixed points are then used in Sec.~\ref{sec:asymptotics} to work out the asymptotic form of charged moments. Finally, Sec.~\ref{sec:conclusions} contains our closing remarks. Some of the technical steps are relegated to the two appendices.

\section{The quantities of interest}\label{sec:quantities}
We assume quantum quench protocol: initially a closed quantum system is
prepared in a state $\ket{\Psi_0}$, and then it is left to evolve unitarily so
that at time $t$ the system is described by a pure state $\ket{\Psi_t}$. We
wish to characterize correlations and fluctuations of conserved charges in the
system, which we will capture by the \emph{charged moments}
$Z_{\alpha,\vec{\beta}}(t)$, defined with respect to an extensive conserved
charge $Q$,
\begin{eqnarray}
  Q=\sum_{j=1}^{2L} q_j, \qquad q_j=\1^{\otimes j-1}\otimes q \otimes \1^{\otimes 2L-j-r+1}
\end{eqnarray}
where $q$ is a local operator that acts nontrivially on $r$ consecutive
sites\footnote{In the general discussion we will always assume $r=1$, as we
implicitly rely on $Q=Q_{A}\oplus Q_{\bar{A}}$. If $r>1$ this does not hold
anymore, but the corrections are subleading in the sizes of subsystems and time
$t$, therefore most of the discussion carries over to finite $r$.}, $2L$ denotes the system size, and we assume periodic boundary conditions. Using $Q$, we define the charged moments as
\begin{eqnarray}\label{eq:genChargedMomentDef} \fl
  Z_{\alpha,\vec{\beta}}(|A|,t)=\tr[
    e^{i \beta_1 Q_A}
    \rho_{A}(t)
    e^{i \beta_2 Q_A}
    \rho_{A}(t)\cdots
    e^{i \beta_{\alpha} Q_A}
    \rho_{A}(t)
  ],\quad
  \vec{\beta}=
  \left[
  \matrix{
    \beta_1 & \beta_2 & \cdots &\beta_{\alpha}
  }
\right],
\end{eqnarray}
where $\alpha\in\mathbb{N}$ is analogous to the R\'enyi index,
$\vec{\beta}\in\mathbb{C}^{\alpha}$, and $\rho_{A}(t)$, $Q_A$ are the state and
charge reduced to a subsystem $A$,
\begin{eqnarray}
  \rho_{A}(t)=\tr_{\bar{A}} \ketbra{\Psi_t}{\Psi_t},\qquad
  Q_{A}=\mkern-14mu
  {\sum_{\scriptstyle j\atop \scriptstyle [j,j+r-1]\subseteq A}}
  \mkern-14mu
  q_j.
\end{eqnarray}

Charged moments contain information about how correlations are shared between
the subsystem $A$ and the complement $\bar{A}$. For instance, in the limit
$\vec{\beta}\to\vec{0}$, 
they encode R\'enyi entanglement entropies, $S_{A}^{(\alpha)}$,
\begin{eqnarray}
  S_{A}^{(\alpha)}(t):=\frac{1}{1-\alpha}
  \log \tr[\rho_{A}^{\alpha}(t)]=
  \frac{1}{1-\alpha}\log Z_{{\alpha},\vec{0}}(|A|,t).
\end{eqnarray}
Another special point is $\alpha\to 1$, $\vec{\beta}\to\beta$, in which case
the charged moments reduce to the \emph{full counting statistics},
\begin{eqnarray}
  Z_{1,\beta}(|A|,t)=\tr[\e^{\beta Q_{A}}\rho_{A}(t)].
\end{eqnarray}

For ${\alpha}>1$, and generic $\vec{\beta}$, the charged moments themselves do not
immediately reduce to known quantities, but they can be related to R\'enyi
entanglement entropies of the state reduced to different symmetry sectors. To
see this, let us first consider a case when the initial state $\ket{\Psi_0}$ 
is an eigenstate of the charge, which means that at time $t$ the state
$\ketbra{\Psi_t}{\Psi_t}$ commutes with $Q$. Then we define $\rho_{A,q}(t)$ as
the reduced state projected to the subspace with the value of charge equal to
$q$,
\begin{eqnarray}\label{eq:defProj}
  \rho_{A,q}(t)=\Pi_{q}\rho_{A}(t) \Pi_{q}=\Pi_{q} \rho_{A}(t),
  \qquad
  \Pi_q=\frac{1}{2\pi}\int_{-\pi}^{\pi} \d\beta \e^{\i \beta(Q_{A}-q)},
\end{eqnarray}
where in the Fourier decomposition of the projector $\Pi_{q}$, we implicitly
assumed $Q$ to have an integer spectrum, and $q\in\mathbb{Z}$. We denote by
$S_{A,q}^{({\alpha})}(t)$ the $\alpha$-R\'enyi entropy of $\rho_{A,q}$,
\begin{eqnarray}
  S_{A,q}^{({\alpha})}(t)=\frac{1}{1-{\alpha}}\tr[\rho_{A,q}^{\alpha}(t)],
\end{eqnarray}
which is referred to as \emph{symmetry resolved R\'enyi entanglement
entropy}~\cite{laflorencie2014spin,goldstein2018symmetry,xavier2018equipartition,bonsignori2019symmetry,murciano2020entanglement}.
Using the above expression for the projector $\Pi_q$, we can express it in
terms of charged moments as~\footnote{The initial state commutes with $Q$,
therefore in the leading order also $\rho_{A}(t)$ commutes with $Q_{A}$, and
$Z_{\alpha,\vec{\beta}}(|A|,t)$ is the same for all $\vec{\beta}$ with the same sum
$\sum_{j=1}^{\alpha}\beta_j$.}
\begin{eqnarray}
  \tr[\rho_{A,q}^{\alpha}(t)]=
  \frac{1}{2\pi}
  \int_{-\pi}^{\pi}\d\beta\, \mathrm{e}^{-\mathrm{i}\beta q}\,
  Z_{{\alpha},[\beta00\ldots0]}(|A|,t).
\end{eqnarray}

On the other hand, if the initial state is \emph{not} an eigenstate of the
charge, an interesting question is how quickly does the reduced density matrix
become block-diagonal in the charge-eigenstate basis. This is measured by the
\emph{entanglement asymmetry}~\cite{ares2023entanglement}, $\Delta S_A(t)$,
which is defined as the relative entropy between the reduced density matrix at time $t$, $\rho_{A}(t)$, and its symmetrized counterpart, $\bar{\rho}_A(t)$,
\begin{eqnarray}
  \Delta S_A(t)=-\tr\left[\rho_A(t)\left(\log\bar{\rho}_A(t)-\log\rho_A(t)\right)\right],
\end{eqnarray}
with 
\begin{eqnarray}
  \bar{\rho}_{A}(t)=\sum_{ q\in\mathbb{Z}} \Pi_q \rho_A(t) \Pi_q.
\end{eqnarray}
The asymmetry can be conveniently expressed as a $\alpha\to1$ limit of the
R\'enyi entanglement asymmetry
\begin{eqnarray}
  \Delta S_A(t)= \lim_{\alpha \to 1} \Delta S^{(\alpha)}(t),\qquad
  \Delta S^{(\alpha)}(t)=\frac{1}{1-\alpha}
  \log\frac{\tr[\bar{\rho}^{\alpha}_A(t)]]}{\tr[\rho^{\alpha}_A(t)]},
\end{eqnarray}
which can in turn be expressed in terms of the charged moments by using  
the Fourier representation of the projector $\Pi_q$ (cf.\ \eref{eq:defProj}),
\begin{eqnarray}
  \tr[\bar{\rho}^{\alpha}_A(t)]=\int_{-\pi}^{\pi}
  \frac{d\beta_1 d\beta_2\cdots d\beta_{\alpha}}{(2\pi)^{\alpha}}
  Z_{\alpha,\vec{\beta}}(|A|,t) \delta(\beta_1+\beta_2+\ldots+\beta_{\alpha}).
\end{eqnarray}
Note that for a charge symmetric initial state, we have
$Z_{\alpha,\vec{\beta}}(|A|,t)=Z_{\alpha,\vec{0}}(|A|,t)$, and therefore $\Delta S^{(\alpha)}(t)=0$.

\section{The setup}\label{sec:setup}
\subsection{The definition of the dynamics}\label{sec:model}
\begin{figure}
  \centering
  \includegraphics[width=0.6\textwidth]{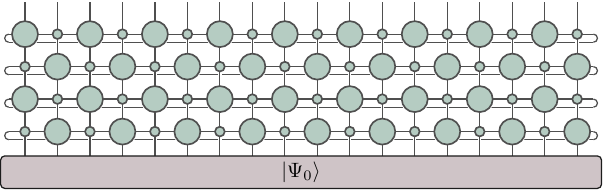}
    \caption{\label{fig:r54TE} Graphical representation of a time-evolved state $\ket{\Psi_t}$
  under the dynamics given by Eq.~\eref{eq:defU} for $L=9$, and $t=2$.}
\end{figure}
The dynamics is defined on a system of $2L$ qubits, with the time-evolution
defined in two discrete time-steps, so that at time $t$ the state of the system
is given as 
\begin{eqnarray} \label{eq:defU}
  \ket{\Psi_t} = ({\mathbb{U}}_\mathrm{o} {\mathbb{U}}_\mathrm{e})^t \ket{\Psi_0},
  \qquad
  {\mathbb{U}}_{\mathrm{e}}=\prod_{j} U_{2j},\qquad
  {\mathbb{U}}_{\mathrm{o}}=\prod_{j} U_{2j-1}.
\end{eqnarray}
Here, $U_{j}$ is an operator that acts as $U$ to the three sites centred around $j$,
\begin{eqnarray}
  U_{j}=\1^{\otimes j-2}\otimes U \otimes \1^{\otimes 2 L-j-1},
\end{eqnarray}
and $U$ is a $3$-site local unitary operator given by the following
computational-basis matrix elements,
\begin{eqnarray}
  \eqalign{
  \mel{s_1^{\prime} s_2^{\prime} s_3^{\prime}}{{U}}{s_1^{\phantom{\prime}} s_2^{\phantom{\prime}} s_3^{\phantom{\prime}}}=
  \delta_{s_1^{\prime},s_1^{\phantom{\prime}}}
  \delta_{s_2^{\prime},\chi(s_1,s_2,s_3)}
  \delta_{s_3^{\prime},s_3^{\phantom{\prime}}},\\
  \chi(s_1,s_2,s_3)\equiv s_1+s_2+s_3+s_1 s_3\pmod{2}.}
\end{eqnarray}
The operator ${U}$ deterministically changes the middle site depending on the
full $3$-site configuration, while the left and right sites are left unchanged.
This implies that all the terms in $\mathbb{U}_{\rm e/o}$ mutually commute and can
be applied at the same time. The time-evolution can therefore be naturally
given in terms of the staggered tensor network pictured in Figure~\ref{fig:r54TE},
where the two circles represent two distinct tensors defined as
\begin{eqnarray} \label{eq:defTETens}
\begin{tikzpicture}[baseline={([yshift=-0.6ex]current bounding box.center)},scale=0.55]
  \gridLine{-0.75}{0}{0.75}{0}
  \gridLine{0}{-0.75}{0}{0.75}
  \sCircle{0}{0}{colU}
  \node at (-1,0) {\scalebox{0.8}{$s_1$}};
  \node at (0,-1) {\scalebox{0.8}{$s_2$}};
  \node at (1,0) {\scalebox{0.8}{$s_3$}};
  \node at (0,1) {\scalebox{0.8}{$s_4$}};
\end{tikzpicture}=\delta_{s_1,s_2}\delta_{s_2,s_3}\delta_{s_3,s_4},\qquad
\begin{tikzpicture}[baseline={([yshift=-0.6ex]current bounding box.center)},scale=0.55]
  \gridLine{-0.75}{0}{0.75}{0}
  \gridLine{0}{-0.75}{0}{0.75}
  \bCircle{0}{0}{colU}
  \node at (-1,0) {\scalebox{0.8}{$s_1$}};
  \node at (0,-1) {\scalebox{0.8}{$s_2$}};
  \node at (1,0) {\scalebox{0.8}{$s_3$}};
  \node at (0,1) {\scalebox{0.8}{$s_4$}};
\end{tikzpicture}=\delta_{s_4,\chi(s_1,s_2,s_3)}.
\end{eqnarray}

\subsection{The simplest conserved charges}
The model is integrable~\cite{friedman2019integrable,gombor2024integrable} and
exhibits an infinite number of conservation laws. Here we will
consider the two conserved quantities with the shortest support,
$Q^{(+)}=\sum_j q^{+}_j$, and $Q^{(-)}=\sum_j q^{-}_j$,
\begin{eqnarray}\fl
  q_j^{(+)}=\frac{1-\sigma_{j+1}^z}{4}
  \left(2+\sigma_{j+2}^z+\sigma_j^z\sigma_{j+2}^z\right)
  ,\qquad
  q_j^{(-)}=\frac{(-1)^j\mkern-8mu}{4}
  \left(1+\sigma_j^z\sigma_{j+1}^z-\sigma_{j}^{z}-\sigma_{j+1}^z\right),
\end{eqnarray}
which can be interpreted as the total number of particles, and their current
respectively~\cite{gopalakrishnan2018hydrodynamics}.
The support of the two operators is larger than $1$, which implies that the object
$e^{i\beta Q^{\pm}_{A}}$ (where $A$ is a finite subsystem) is not in a product form
(contrary to $U(1)$ charges considered in~\cite{bertini2023nonequilibrium}). However, their finite support implies that they exhibit an efficient representation as a matrix-product-operator (MPO), see e.g.\ \cite{haegeman2017diagonalizing}, with a finite bond dimension.
In particular, in the case of $Q^{(-)}$, the auxiliary space is two-dimensional,
and the MPO has the following staggered structure,
\begin{eqnarray}\label{eq:MPOqmDef}
  e^{i\beta Q^{-}_A}=
  \begin{tikzpicture}[baseline={([yshift=-0.6ex]current bounding box.center)},scale=0.6]
    \def\X{6}
    \vLine{0.5}{0}{\X+0.5}{0}
    \vBmps{0.5}{0}{colObsLines}
    \vBmps{\X+0.5}{0}{colObsLines}
    \foreach \x in {1,3,...,\X}{
      \gridLine{\x}{-0.5}{\x}{0.5}
      \gridLine{\x+1}{-0.5}{\x+1}{0.5}
      \vmpsV{\x}{0}{colObs}{colObsLines}
      \vmpsW{\x+1}{0}{colObs}{colObsLines}
    }
  \end{tikzpicture},
\end{eqnarray}
where the vertical legs represent $1$-qubit physical degrees of freedom, while the two
sets of matrices are
\begin{eqnarray}
  \begin{tikzpicture}[baseline={([yshift=-0.6ex]current bounding box.center)},scale=0.6]
    \vLine{-0.5}{0}{0.5}{0}
    \gridLine{0}{-0.5}{0}{0.5}
    \vmpsV{0}{0}{colObs}{colObsLines}
    \node at (0,0.75) {\scalebox{0.8}{$s$}};
    \node at (0,-0.75) {\scalebox{0.8}{$b$}};
  \end{tikzpicture}=
  \delta_{s,0}\delta_{b,0}
  \left[\matrix{ 1 & 0 \cr 1 & 0 }\right]
  +\delta_{s,1}\delta_{b,1}
  \left[\matrix{0 & 1 \cr 0 & e^{i \beta}}\right]
  ,\qquad
  \begin{tikzpicture}[baseline={([yshift=-0.6ex]current bounding box.center)},scale=0.6]
    \vLine{-0.5}{0}{0.5}{0}
    \gridLine{0}{-0.5}{0}{0.5}
    \vmpsW{0}{0}{colObs}{colObsLines}
  \end{tikzpicture}=
  \left.
  \begin{tikzpicture}[baseline={([yshift=-0.6ex]current bounding box.center)},scale=0.6]
    \vLine{-0.5}{0}{0.5}{0}
    \gridLine{0}{-0.5}{0}{0.5}
    \vmpsV{0}{0}{colObs}{colObsLines}
  \end{tikzpicture}
  \right|_{\beta\leftrightarrow-\beta},
\end{eqnarray}
and the boundary vectors in the auxiliary space are chosen as
\begin{eqnarray} \label{eq:boundaryMPO}
  \begin{tikzpicture}[baseline={([yshift=-0.6ex]current bounding box.center)},scale=0.6]
    \vLine{-0.5}{0}{0}{0}
    \vBmps{-0.5}{0}{colObsLines}
    \end{tikzpicture}=
    \left.\left[\matrix{1\cr 0}\right]\mkern-4mu\right.^T,\qquad
  \begin{tikzpicture}[baseline={([yshift=-0.6ex]current bounding box.center)},scale=0.6]
    \vLine{0.5}{0}{0}{0}
    \vBmps{0.5}{0}{colObsLines}
  \end{tikzpicture}=\left[\matrix{1\cr 1}\right].
\end{eqnarray}

Similarly, $e^{i\beta Q^{(+)}}$ can be also represented as an MPO,
but the larger support implies a larger bond dimension (in particular, it is $3$),
and there is no even-odd staggering effect,
\begin{eqnarray}\label{eq:MPOqpDef}
  e^{i\beta Q^{+}_A}=
  \begin{tikzpicture}[baseline={([yshift=-0.6ex]current bounding box.center)},scale=0.6]
    \def\X{6}
    \vLine{0.5}{0}{\X+0.5}{0}
    \vBmps{0.5}{0}{colObsLines}
    \vBmps{\X+0.5}{0}{colObsLines}
    \foreach \x in {1,3,...,\X}{
      \gridLine{\x}{-0.5}{\x}{0.5}
      \gridLine{\x+1}{-0.5}{\x+1}{0.5}
      \vmpsNS{\x}{0}{colObs}{colObsLines}
      \vmpsNS{\x+1}{0}{colObs}{colObsLines}
    }
  \end{tikzpicture},
\end{eqnarray}
with 
\begin{eqnarray}\fl
  \begin{tikzpicture}[baseline={([yshift=-0.6ex]current bounding box.center)},scale=0.6]
    \vLine{-0.5}{0}{0.5}{0}
    \gridLine{0}{-0.5}{0}{0.5}
    \vmpsNS{0}{0}{colObs}{colObsLines}
    \node at (0,0.75) {\scalebox{0.8}{$s$}};
    \node at (0,-0.75) {\scalebox{0.8}{$b$}};
  \end{tikzpicture}=
  \delta_{s,0}\delta_{b,0}\left[\matrix{1&0&0\cr e^{2 i \beta}&0&0\cr 1&0&0}\right]
  +\delta_{s,1}\delta_{b,1}\left[\matrix{0&1&0\cr 0&0&e^{i \beta}\cr 0&0&e^{i\beta}}\right]
  \mkern-6mu,\ 
  \begin{tikzpicture}[baseline={([yshift=-0.6ex]current bounding box.center)},scale=0.6]
    \vLine{-0.5}{0}{0}{0}
    \vBmps{-0.5}{0}{colObsLines}
  \end{tikzpicture}=
  \left.\left[\matrix{0 \cr \frac{e^{-i\beta}}{1+e^{i\beta}} \cr 
  \frac{1}{1+e^{i\beta}}}\right]\mkern-4mu\right.^{T}
  \mkern-14mu,\ 
  \begin{tikzpicture}[baseline={([yshift=-0.6ex]current bounding box.center)},scale=0.6]
    \vLine{0.5}{0}{0}{0}
    \vBmps{0.5}{0}{colObsLines}
  \end{tikzpicture}=
  \left[\matrix{
    1\cr 
  1\cr 
  e^{-i\beta}
  }\right]\mkern-6mu.
\end{eqnarray}
Note that for simplicity we use the same notation for the auxiliary degrees of freedom
and boundary vectors in the two cases, but it is always clear from the context which one
is used.

\subsection{Initial states}
The reduced density matrix $\rho_A(t)$ is the state $\rho(t)=\ketbra{\Psi_t}{\Psi_t}$
reduced to the subsystem $A$, where the initial state $\ket{\Psi_0}$ is chosen to be
from the family of \emph{solvable} initial states introduced
in~\cite{klobas2021exact,klobas2021exactrelaxation},
\begin{eqnarray}\label{eq:defSolvableIS}
  \fl
    \ket{\Psi_0}=\left(\ket{\psi_1}\otimes \ket{\psi_2}\right)^{\otimes L}\mkern-8mu,\quad
  \ket{\psi_1}=
  \begin{tikzpicture}[baseline={([yshift=-0.6ex]current bounding box.center)},scale=0.55]
    \gridLine{1}{0}{1}{0.75}
    \vmpsV{1}{0}{colIst}{colLines}
  \end{tikzpicture}=
  \left[\matrix{1\cr 0 }\right]\mkern-4mu,\quad
  \ket{\psi_2}=
  \begin{tikzpicture}[baseline={([yshift=-0.6ex]current bounding box.center)},scale=0.55]
    \gridLine{1}{0}{1}{0.75}
    \vmpsW{1}{0}{colIst}{colLines}
  \end{tikzpicture}
  =\left[\matrix{\sqrt{1-\vartheta}\cr
  \sqrt{\vartheta}}\right]\mkern-4mu,\quad
  0<\vartheta<1.
\end{eqnarray}
After being initialized in one of these states, the system can be
shown~\cite{klobas2021exactrelaxation} to relax to the family of Gibbs states
determined by the first one of the conserved quantities introduced above,
$\rho\sim e^{\mu Q^{(+)}}$, with the chemical potential $\mu=\log(\vartheta^{-1}-1)$.

\subsection{Space evolution}
\begin{figure}
  \centering
  \includegraphics[width=0.6\textwidth]{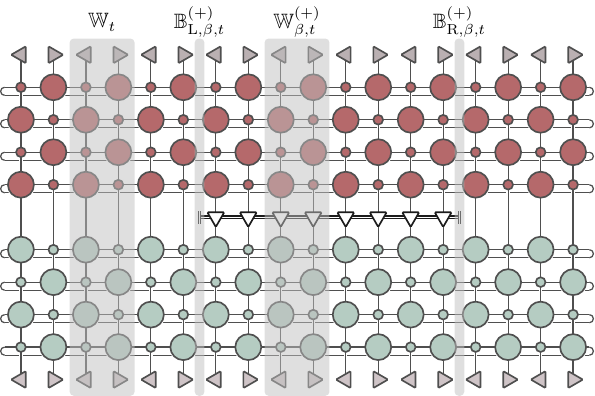}
  \caption{\label{fig:Z1beta}
  The diagrammatic representation of $Z_{1,\beta}^{+}(l,t)$ for $L=9$, $l=4$, and $t=2$. The shaed boxes highlight the objects appearing in Eq.~\eref{eq:exampleZ1beta}.}
\end{figure}

The definitions above completely specify our setup, and we are now able to consider the quantities $Z^{(\pm)}_{\alpha,\vec{\beta}}(l,t)$, given as charged moments (cf.\ \eref{eq:genChargedMomentDef}), where the label $\pm$ refers to either $Q^{(+)}$, or $Q^{(-)}$ in the definition of the charged moment, the initial state is given by~\eref{eq:defSolvableIS}, 
and the subsystem $A$ has the length $|A|=l$. Note that we define the length $|A|$ of the subsystem $A$ to coincide with half of the number of sites in $A$ (i.e.\ the number of sites in $A$ is $2|A|$). 
By denoting the Hermitian conjugates of time-evolution tensors~\eref{eq:defTETens} with red,\footnote{We note that in our case the red and green tensors are the same, as the model is left-right symmetric, deterministic, and reversible. However, the formulation with two different colours naturally generalises to a more general setting, therefore we decided to keep the distinction between the two sets of tensors.}
\begin{eqnarray}
\begin{tikzpicture}[baseline={([yshift=-0.6ex]current bounding box.center)},scale=0.55]
  \gridLine{-0.75}{0}{0.75}{0}
  \gridLine{0}{-0.75}{0}{0.75}
  \sCircle{0}{0}{colUc}
\end{tikzpicture}=
\left.
  \begin{tikzpicture}[baseline={([yshift=-0.6ex]current bounding box.center)},scale=0.55]
    \gridLine{-0.75}{0}{0.75}{0}
    \gridLine{0}{-0.75}{0}{0.75}
    \sCircle{0}{0}{colU}
  \end{tikzpicture}
\right.^{\dagger},\qquad
\begin{tikzpicture}[baseline={([yshift=-0.6ex]current bounding box.center)},scale=0.55]
  \gridLine{-0.75}{0}{0.75}{0}
  \gridLine{0}{-0.75}{0}{0.75}
  \bCircle{0}{0}{colUc}
\end{tikzpicture}=
\left.
  \begin{tikzpicture}[baseline={([yshift=-0.6ex]current bounding box.center)},scale=0.55]
    \gridLine{-0.75}{0}{0.75}{0}
    \gridLine{0}{-0.75}{0}{0.75}
    \bCircle{0}{0}{colU}
  \end{tikzpicture}
\right.^{\dagger},
\end{eqnarray}
we are able to represent e.g.\ $Z_{1,\beta}^{(+)}(l,t)$ as shown in Fig.~\ref{fig:Z1beta}.
The graphical representation of the tensor network suggests that $Z_{1,\beta}^{(+)}$ can
be equivalently contracted in \emph{space} rather than in time. In particular, we introduce
Hilbert space of the \emph{temporal chain}  of $2t$ qubits,
\begin{eqnarray}
  \mathcal{V}_{t}=\left(\mathbb{C}^2\right)^{\otimes 2t},
\end{eqnarray}
and define transfer matrices $\mathbb{W}_t$, $\mathbb{W}_{\beta,t}^{(\pm)}$ that act on
$2\times 2t$ vertical lattice sites, and (possibly) the auxiliary degree of freedom, 
\begin{eqnarray}\fl
  \mathbb{W}_t\in\End(\mathcal{V}_t\otimes \mathcal{V}_t),\quad
  \mathbb{W}_{\beta,t}^{(+)}\in
  \End(\mathcal{V}_t\otimes \mathbb{C}^{3}\otimes \mathcal{V}_t),\quad
 \mathbb{W}_{\beta,t}^{(-)}\in
  \End(\mathcal{V}_t\otimes \mathbb{C}^{2}\otimes \mathcal{V}_t).
\end{eqnarray}
In particular, space transfer matrices are built by two consecutive rows of 
the diagram in Fig.~\ref{fig:Z1beta} (and an analogous one corresponding to
$Z_{1,\beta}^{(-)}(l,t)$),
\begin{eqnarray}
  \mathbb{W}_t=
  \begin{tikzpicture}[baseline={([yshift=-0.6ex]current bounding box.center)},scale=0.55]
    \def\Y{4}

    \gridLine{1}{0}{1}{2*\Y+2}
    \gridLine{2}{0}{2}{2*\Y+2}

    \foreach \t in {1,2,...,\Y}
    {
      \gridLine{0.25}{\t}{2.75}{\t}
      \gridLine{0.25}{\Y+1+\t}{2.75}{\Y+1+\t}
    }
    \foreach \t in {1,3,...,\Y}
    {
      \sCircle{1}{\t}{colU}
      \bCircle{2}{\t}{colU}
      \bCircle{1}{\t+1}{colU}
      \sCircle{2}{\t+1}{colU}

      \bCircle{1}{\t+\Y+1}{colUc}
      \sCircle{2}{\t+\Y+1}{colUc}
      \sCircle{1}{\t+\Y+2}{colUc}
      \bCircle{2}{\t+\Y+2}{colUc}
    }
    \vmpsV{1}{0}{colIst}{colLines}
    \vmpsW{2}{0}{colIst}{colLines}
    \vmpsV{1}{2*\Y+2}{colIstC}{colLines}
    \vmpsW{2}{2*\Y+2}{colIstC}{colLines}
  \end{tikzpicture},\qquad
  \mathbb{W}_{\beta,t}^{(+)}=
  \begin{tikzpicture}[baseline={([yshift=-0.6ex]current bounding box.center)},scale=0.55]
    \def\Y{4}

    \gridLine{1}{0}{1}{2*\Y+2}
    \gridLine{2}{0}{2}{2*\Y+2}

    \foreach \t in {1,2,...,\Y}
    {
      \gridLine{0.25}{\t}{2.75}{\t}
      \gridLine{0.25}{\Y+1+\t}{2.75}{\Y+1+\t}
    }
    \foreach \t in {1,3,...,\Y}
    {
      \sCircle{1}{\t}{colU}
      \bCircle{2}{\t}{colU}
      \bCircle{1}{\t+1}{colU}
      \sCircle{2}{\t+1}{colU}

      \bCircle{1}{\t+\Y+1}{colUc}
      \sCircle{2}{\t+\Y+1}{colUc}
      \sCircle{1}{\t+\Y+2}{colUc}
      \bCircle{2}{\t+\Y+2}{colUc}
    }
    \vmpsV{1}{0}{colIst}{colLines}
    \vmpsW{2}{0}{colIst}{colLines}
    \vmpsV{1}{2*\Y+2}{colIstC}{colLines}
    \vmpsW{2}{2*\Y+2}{colIstC}{colLines}

    \vLine{0.25}{\Y+1}{2.75}{\Y+1}
    \vmpsNS{1}{\Y+1}{colObs}{colObsLines}
    \vmpsNS{2}{\Y+1}{colObs}{colObsLines}
  \end{tikzpicture},\qquad
  \mathbb{W}_{\beta,t}^{(-)}=
  \begin{tikzpicture}[baseline={([yshift=-0.6ex]current bounding box.center)},scale=0.55]
    \def\Y{4}

    \gridLine{1}{0}{1}{2*\Y+2}
    \gridLine{2}{0}{2}{2*\Y+2}

    \foreach \t in {1,2,...,\Y}
    {
      \gridLine{0.25}{\t}{2.75}{\t}
      \gridLine{0.25}{\Y+1+\t}{2.75}{\Y+1+\t}
    }
    \foreach \t in {1,3,...,\Y}
    {
      \sCircle{1}{\t}{colU}
      \bCircle{2}{\t}{colU}
      \bCircle{1}{\t+1}{colU}
      \sCircle{2}{\t+1}{colU}

      \bCircle{1}{\t+\Y+1}{colUc}
      \sCircle{2}{\t+\Y+1}{colUc}
      \sCircle{1}{\t+\Y+2}{colUc}
      \bCircle{2}{\t+\Y+2}{colUc}
    }
    \vmpsV{1}{0}{colIst}{colLines}
    \vmpsW{2}{0}{colIst}{colLines}
    \vmpsV{1}{2*\Y+2}{colIstC}{colLines}
    \vmpsW{2}{2*\Y+2}{colIstC}{colLines}

    \vLine{0.25}{\Y+1}{2.75}{\Y+1}
    \vmpsV{1}{\Y+1}{colObs}{colObsLines}
    \vmpsW{2}{\Y+1}{colObs}{colObsLines}
  \end{tikzpicture}.
\end{eqnarray}
In terms of these transfer matrices the moments 
$Z_{1,\beta}^{(\pm)}(l,t)$ take the following form,
\begin{eqnarray}\label{eq:exampleZ1beta}
  Z_{1,\beta}^{(\pm)}(l,t)=
  \tr\Big[ 
  \mathbb{W}^{L-l}_t
  \mathbb{B}_{{\rm L},\beta,t}^{(\pm)}
  \left.
  \mathbb{W}_{\beta,t}^{(\pm)}\right.^{l}
  \mathbb{B}_{{\rm R},\beta,t}^{(\pm)}
  \Big],
\end{eqnarray}
where $\mathbb{B}_{{\rm L/R},\beta,t}^{(\pm)}$ act as identity on both $\mathbb{V}_t$,
and as boundary vectors
$\begin{tikzpicture}[baseline={([yshift=-0.6ex]current bounding box.center)},scale=0.5]
    \vLine{0.5}{0}{0}{0}
    \vBmps{0.5}{0}{colObsLines}
\end{tikzpicture}$,
$\begin{tikzpicture}[baseline={([yshift=-0.6ex]current bounding box.center)},scale=0.5]
    \vLine{-0.5}{0}{0}{0}
    \vBmps{-0.5}{0}{colObsLines}
\end{tikzpicture}$
on the auxiliary degree of freedom (see Fig.~\ref{fig:Z1beta}).

More complicated charged moments $Z_{\alpha,\vec{\beta}}^{(\pm)}(l,t)$ are obtained
by considering $\alpha$ copies of transfer matrices that are connected in a staggered way,
\begin{eqnarray}\fl \label{eq:CMspace}
  Z_{\alpha,\vec{\beta}}^{(\pm)}(l,t)=
  \tr\Big[ 
  \mathbb{P}_{\alpha,t}^{\dagger}
  \left(\mathbb{W}_t^{\otimes \alpha}\right)^{L-l}
  \mathbb{P}_{\alpha,t}
  \mathbb{B}_{{\rm L},\alpha,\vec{\beta},t}^{(\pm)}
  \left(
  \mathbb{W}_{\beta_1,t}^{(\pm)}\otimes
  \mathbb{W}_{\beta_2,t}^{(\pm)}\otimes
  \cdots
  \otimes
  \mathbb{W}_{\beta_{\alpha},t}^{(\pm)}\right)^{l}
  \mathbb{B}_{{\rm R},\alpha,\vec{\beta},t}^{(\pm)}\Big],
\end{eqnarray}
where $\mathbb{B}_{{\rm L/R},\alpha,\vec{\beta},t}^{(\pm)}$ is the
multi-copy generalization of $\mathbb{B}_{{\rm L/R},\beta,t}^{(\pm)}$,
\begin{eqnarray}
  \mathbb{B}_{{\rm L/R},\alpha,\vec{\beta},t}^{(\pm)}=
  \mathbb{B}_{{\rm L/R},\beta_1,t}^{(\pm)}\otimes
  \mathbb{B}_{{\rm L/R},\beta_2,t}^{(\pm)}\otimes\cdots\otimes
  \mathbb{B}_{{\rm L/R},\beta_{\alpha},t}^{(\pm)},
\end{eqnarray}
and $\mathbb{P}_{\alpha,t}\in\End(\mathcal{V}_{t}^{\otimes 2\alpha})$ 
is a permutation operator that periodically shifts \emph{even} vectors in
a tensor product of $2\alpha$ states from $\mathcal{V}_t$,
\begin{eqnarray}\fl
  \mathbb{P}_{\alpha,t} 
  \ket{x_1}\otimes \ket{x_2}\otimes \cdots \otimes \ket{x_{2 \alpha}} =
  \ket{x_1}\otimes \ket{x_{2\alpha}}\otimes \ket{x_3}\otimes \ket{x_2} \otimes \cdots
  \otimes \ket{x_{2\alpha-2}}.
\end{eqnarray}

\section{Fixed-points of the space transfer matrix}\label{sec:FPs}
The representation of charged moments in terms of space transfer matrices (cf.\
\eref{eq:CMspace}) in general offers no benefit as it is completely equivalent
to the definition~\eref{eq:genChargedMomentDef}.  However, the locality and
unitarity of time-evolution imply that $\mathbb{W}_t$ has a unique maximal
eigenvalue $1$, all the other eigenvalues are equal to $0$, and the largest
Jordan block is of size at most $2t+1$. Therefore, a high-enough power of
$\mathbb{W}_t$ can be substituted by a projector to its \emph{fixed points}
(i.e.\ leading left and right eigenvectors),
\begin{eqnarray}\label{eq:SpectW}
  \mathbb{W}_t \ket{r_{t}}=\ket{r_t},\qquad
  \bra{l_t}\mathbb{W}_t=\bra{l_t},\qquad
  \left.\mathbb{W}_t^x\right|_{x>2t+1}=\ketbra{r_t}{l_t}.
\end{eqnarray}
The additional auxiliary degree of freedom due to $e^{i\beta Q^{(\pm)}_A}$
changes the spectral properties of the transfer matrix
$\mathbb{W}_{\beta,t}^{(\pm)}$, so that the leading eigenvalue is no longer
necessarily $1$. Nonetheless, as is shown in~\ref{sec:uniquenessFPs}, the
leading eigenvalue $\Lambda_{\beta}^{(\pm)}$ is $t$-independent and unique, and
a repeated application of the transfer matrix again projects to the leading
eigenvectors, 
\begin{eqnarray}\label{eq:SpectWbeta}
  \eqalign{
    \mathbb{W}_{\beta,t}^{(\pm)} \sket{r_{\beta,t}^{(\pm)}}=\Lambda_{\beta}^{(\pm)}
    \sket{r_{\beta,t}^{(\pm)}},\qquad
    \sbra{l_{\beta,t}^{(\pm)}}\mathbb{W}_{\beta,t}^{(\pm)}=\Lambda_{\beta}^{(\pm)}
    \sbra{l_{\beta,t}^{(\pm)}},\\
    \lim_{x\to\infty} 
    \Big(\frac{1}{\Lambda_{\beta}^{(\pm)}}\mathbb{W}_{\beta,t}^{(\pm)} \Big)^x
    =\frac{\sketbra{r_{\beta,t}^{(\pm)}}{l_{\beta,t}^{(\pm)}}}
    {\sbraket{r_{\beta,t}^{(\pm)}}{l_{\beta,t}^{(\pm)}}}.}
\end{eqnarray}
Therefore in the regime of large system and subsystem sizes (compared to $t$)
the behaviour of charged moments $Z_{\alpha,\vec{\beta}}^{(\pm)}(l,t)$ is
completely determined by the properties of leading eigenvalues $\Lambda_{\beta}^{(\pm)}$,
and the corresponding eigenvectors
$\sket{r_{\beta,t}^{(\pm)}}$, $\sbra{l_{\beta,t}^{(\pm)}}$, $\sket{r_t}$, 
$\sbra{l_t}$.

\subsection{Fixed points of $\mathbb{W}_t$}
We start by summarizing the construction of the fixed points of $\mathbb{W}_t$,
which was introduced in Refs.~\cite{klobas2021exact,klobas2021exactrelaxation}.
The fixed points can be conveniently expressed in the following MPO form,
\begin{eqnarray}\label{eq:defAnstazFPmpo}
  \bra{l_t}=
  \begin{tikzpicture}[baseline={([yshift=-0.6ex]current bounding box.center)},scale=0.55]
    \draw[thick,colLines,rounded corners] (0.75,4) -- (-0.75,4) -- (-0.75,6) -- (0.75,6);
    \draw[thick,colLines,rounded corners] (0.75,3) -- (-0.875,3) -- (-0.875,7) -- (0.75,7);
    \draw[thick,colLines,rounded corners] (0.75,2) -- (-1,2) -- (-1,8) -- (0.75,8);
    \draw[thick,colLines,rounded corners] (0.75,1) -- (-1.125,1) -- (-1.125,9) -- (0.75,9);
    \mpsWire{0}{0}{0}{5}
    \mpsBvecW{0}{0}{colMPS}
    \mpsA{0}{1}{colMPS}
    \mpsB{0}{2}{colMPS}
    \mpsA{0}{3}{colMPS}
    \mpsB{0}{4}{colMPS}
    \mpsBvec{0}{5}
  \end{tikzpicture},\qquad
  \ket{r_t}=
  \begin{tikzpicture}[baseline={([yshift=-0.6ex]current bounding box.center)},scale=0.55]
    \draw[thick,colLines,rounded corners] (-0.75,4) -- (0.75,4) -- (0.75,6) -- (-0.75,6);
    \draw[thick,colLines,rounded corners] (-0.75,3) -- (0.875,3) -- (0.875,7) -- (-0.75,7);
    \draw[thick,colLines,rounded corners] (-0.75,2) -- (1,2) -- (1,8) -- (-0.75,8);
    \draw[thick,colLines,rounded corners] (-0.75,1) -- (1.125,1) -- (1.125,9) -- (-0.75,9);
    \mpsWire{0}{0}{0}{5}
    \mpsBvecV{0}{0}{colMPS}
    \mpsB{0}{1}{colMPS}
    \mpsA{0}{2}{colMPS}
    \mpsB{0}{3}{colMPS}
    \mpsA{0}{4}{colMPS}
    \mpsBvec{0}{5}
  \end{tikzpicture},
\end{eqnarray}
where the ticker vertical line represents the auxiliary vector space. It is
straightforward to see that the ansatz for $\bra{l_t}$ gives the left fixed
point of the transfer matrix, if the following set of algebraic relations is
fulfilled,
\begin{eqnarray}\label{eq:localRelationsLeftBulkBottom}
  \begin{tikzpicture}[baseline={([yshift=0.7ex]current bounding box.center)},scale=0.5]
    \mpsWire{0}{0.25}{0}{4.75}
    \draw[thick,colLines,rounded corners](1.75,4) -- (-0.5,4) -- (-0.5,5.5) -- (1.75,5.5);
    \draw[thick,colLines,rounded corners](1.75,3) -- (-0.625,3) -- (-0.625,6.5) -- (1.75,6.5);
    \draw[thick,colLines,rounded corners](1.75,2) -- (-0.75,2) -- (-0.75,7.5) -- (1.75,7.5);
    \draw[thick,colLines,rounded corners](1.75,1) -- (-0.875,1) -- (-0.875,8.5) -- (1.75,8.5);
    \mpsC{0}{1}{2}{colMPS}
    \mpsB{0}{3}{colMPS}
    \mpsA{0}{4}{colMPS}
    \gridLine{1}{2}{1}{4}
    \gridLine{1}{5.5}{1}{7.5}
    \sCircle{1}{2}{colU}
    \bCircle{1}{3}{colU}
    \sCircle{1}{4}{colU}
    \sCircle{1}{5.5}{colUc}
    \bCircle{1}{6.5}{colUc}
    \sCircle{1}{7.5}{colUc}
  \end{tikzpicture}=
  \begin{tikzpicture}[baseline={([yshift=0.7ex]current bounding box.center)},scale=0.5]
    \mpsWire{0}{0.25}{0}{4.75}
    \draw[thick,colLines,rounded corners]
    (0.75,4) -- (-0.5,4) -- (-0.5,5.5) -- (0.75,5.5);
    \draw[thick,colLines,rounded corners]
    (0.75,3) -- (-0.625,3) -- (-0.625,6.5) -- (0.75,6.5);
    \draw[thick,colLines,rounded corners]
    (0.75,2) -- (-0.75,2) -- (-0.75,7.5) -- (0.75,7.5);
    \draw[thick,colLines,rounded corners]
    (0.75,1) -- (-0.875,1) -- (-0.875,8.5) -- (0.75,8.5);
    \mpsC{0}{3}{4}{colMPS}
    \mpsA{0}{2}{colMPS}
    \mpsB{0}{1}{colMPS}
  \end{tikzpicture},\qquad
  \begin{tikzpicture}[baseline={([yshift=-0.6ex]current bounding box.center)},scale=0.5]
    \mpsWire{0}{0}{0}{2.75}
    \draw[thick,colLines,rounded corners] (1.75,2) -- (-0.5,2) -- (-0.5,3.5) -- (1.75,3.5);
    \draw[thick,colLines,rounded corners] (1.75,1) -- (-0.625,1) -- (-0.625,4.5) -- (1.75,4.5);
    \mpsBvecV{0}{0}{colMPS}
    \mpsB{0}{1}{colMPS}
    \mpsA{0}{2}{colMPS}
    \gridLine{1}{0}{1}{2}
    \gridLine{1}{3.5}{1}{5.5}
    \sCircle{1}{2}{colU}
    \bCircle{1}{1}{colU}
    \sCircle{1}{3.5}{colUc}
    \bCircle{1}{4.5}{colUc}
    \vmpsW{1}{0}{colIst}{colLines}
    \vmpsW{1}{5.5}{colIstC}{colLines}
  \end{tikzpicture}=
  \begin{tikzpicture}[baseline={([yshift=1.5ex]current bounding box.center)},scale=0.5]
    \mpsWire{0}{0}{0}{2.75}
    \draw[thick,colLines,rounded corners] (0.75,2) -- (-0.5,2) -- (-0.5,3.5) -- (0.75,3.5);
    \draw[thick,colLines,rounded corners] (0.75,1) -- (-0.625,1) -- (-0.625,4.5) -- (0.75,4.5);
    \mpsBvecW{0}{0}{colMPS}
    \mpsC{0}{1}{2}{colMPS}
  \end{tikzpicture},\qquad
  \begin{tikzpicture}[baseline={([yshift=-0.6ex]current bounding box.center)},scale=0.5]
    \mpsWire{0}{0}{0}{3.75}
    \draw[thick,colLines,rounded corners] (1.75,3) -- (-0.5,3) -- (-0.5,4.5) -- (1.75,4.5);
    \draw[thick,colLines,rounded corners] (1.75,2) -- (-0.625,2) -- (-0.625,5.5) -- (1.75,5.5);
    \draw[thick,colLines,rounded corners] (1.75,1) -- (-0.75,1) -- (-0.75,6.5) -- (1.75,6.5);
    \mpsBvecW{0}{0}{colMPS}
    \mpsA{0}{1}{colMPS}
    \mpsB{0}{2}{colMPS}
    \mpsA{0}{3}{colMPS}
    \gridLine{1}{0}{1}{3}
    \gridLine{1}{4.5}{1}{7.5}
    \sCircle{1}{3}{colU}
    \bCircle{1}{2}{colU}
    \sCircle{1}{1}{colU}
    \sCircle{1}{4.5}{colUc}
    \bCircle{1}{5.5}{colUc}
    \sCircle{1}{6.5}{colUc}
    \vmpsV{1}{0}{colIst}{colLines}
    \vmpsV{1}{7.5}{colIstC}{colLines}
  \end{tikzpicture}=
  \begin{tikzpicture}[baseline={([yshift=1.5ex]current bounding box.center)},scale=0.5]
    \mpsWire{0}{0}{0}{3.75}
    \draw[thick,colLines,rounded corners] (0.75,3) -- (-0.5,3) -- (-0.5,4.5) -- (0.75,4.5);
    \draw[thick,colLines,rounded corners] (0.75,2) -- (-0.625,2) -- (-0.625,5.5) -- (0.75,5.5);
    \draw[thick,colLines,rounded corners] (0.75,1) -- (-0.75,1) -- (-0.75,6.5) -- (0.75,6.5);
    \mpsBvecV{0}{0}{colMPS}
    \mpsB{0}{1}{colMPS}
    \mpsC{0}{2}{3}{colMPS}
  \end{tikzpicture},\\
  \label{eq:localRelationsLeftTop}
  \begin{tikzpicture}[baseline={([yshift=0.7ex]current bounding box.center)},scale=0.5]
    \mpsWire{0}{0.25}{0}{4}
    \draw[thick,colLines,rounded corners] (1.75,3) -- (-0.5,3) -- (-0.5,5) -- (1.75,5);
    \draw[thick,colLines,rounded corners] (1.75,2) -- (-0.625,2) -- (-0.625,6) -- (1.75,6);
    \draw[thick,colLines,rounded corners] (1.75,1) -- (-0.75,1) -- (-0.75,7) -- (1.75,7);
    \mpsBvec{0}{4}
    \mpsC{0}{1}{2}{colMPS}
    \mpsB{0}{3}{colMPS}
    \gridLine{1}{2}{1}{6}
    \sCircle{1}{2}{colU}
    \bCircle{1}{3}{colU}
    \bCircle{1}{5}{colUc}
    \sCircle{1}{6}{colUc}
  \end{tikzpicture}=
  \begin{tikzpicture}[baseline={([yshift=0.7ex]current bounding box.center)},scale=0.5]
    \mpsWire{0}{0.25}{0}{4}
    \draw[thick,colLines,rounded corners] (0.75,3) -- (-0.5,3) -- (-0.5,5) -- (0.75,5);
    \draw[thick,colLines,rounded corners] (0.75,2) -- (-0.625,2) -- (-0.625,6) -- (0.75,6);
    \draw[thick,colLines,rounded corners] (0.75,1) -- (-0.75,1) -- (-0.75,7) -- (0.75,7);
    \mpsBvec{0}{4}
    \mpsA{0}{1}{colMPS}
    \mpsB{0}{2}{colMPS}
    \mpsA{0}{3}{colMPS}
  \end{tikzpicture},\qquad
  \begin{tikzpicture}[baseline={([yshift=0.7ex]current bounding box.center)},scale=0.5]
    \mpsWire{0}{0.25}{0}{3}
    \draw[thick,colLines,rounded corners] (1.75,2) -- (-0.5,2) -- (-0.5,4) -- (1.75,4);
    \draw[thick,colLines,rounded corners] (1.75,1) -- (-0.625,1) -- (-0.625,5) -- (1.75,5);
    \mpsBvec{0}{3}
    \mpsC{0}{1}{2}{colMPS}
    \gridLine{1}{2}{1}{4}
    \sCircle{1}{2}{colU}
    \sCircle{1}{4}{colUc}
  \end{tikzpicture}=
  \begin{tikzpicture}[baseline={([yshift=0.7ex]current bounding box.center)},scale=0.5]
    \mpsWire{0}{0.25}{0}{3}
    \draw[thick,colLines,rounded corners] (0.75,2) -- (-0.5,2) -- (-0.5,4) -- (0.75,4);
    \draw[thick,colLines,rounded corners] (0.75,1) -- (-0.625,1) -- (-0.625,5) -- (0.75,5);
    \mpsBvec{0}{3}
    \mpsB{0}{2}{colMPS}
    \mpsA{0}{1}{colMPS}
  \end{tikzpicture},
\end{eqnarray}
and an analogous set of left-right flipped relations ensures that $\ket{r_t}$ is
the right fixed point. A solution to these relations exists for the $3$-dimensional
auxiliary vector space~\cite{klobas2021exactrelaxation},
\begin{eqnarray}\label{eq:originalFixedPointTensors}\fl
  \eqalign{
    &\begin{tikzpicture}[baseline={([yshift=-0.6ex]current bounding box.center)},scale=0.5]
      \mpsWire{0}{-0.625}{0}{0.625}
      \gridLine{-0.55}{0}{0.55}{0}
      \node at (-0.75,0) {\scalebox{0.8}{$0$}};
      \node at (0.75,0) {\scalebox{0.8}{$0$}};
      \mpsA{0}{0}{colMPS}
    \end{tikzpicture}=
    \left[\matrix{
      1-\vartheta&1-\vartheta&-1+\vartheta\cr
      \vartheta&\vartheta&1-\vartheta\cr
      \vartheta&-\frac{\vartheta^2}{1-\vartheta}&-\vartheta
    }\right],\\
    &\begin{tikzpicture}[baseline={([yshift=-0.6ex]current bounding box.center)},scale=0.5]
      \mpsWire{0}{-0.625}{0}{0.625}
      \gridLine{-0.55}{0}{0.55}{0}
      \node at (-0.75,0) {\scalebox{0.8}{$0$}};
      \node at (0.75,0) {\scalebox{0.8}{$1$}};
      \mpsA{0}{0}{colMPS}
    \end{tikzpicture}=
    \begin{tikzpicture}[baseline={([yshift=-0.6ex]current bounding box.center)},scale=0.5]
      \mpsWire{0}{-0.625}{0}{0.625}
      \gridLine{-0.55}{0}{0.55}{0}
      \node at (-0.75,0) {\scalebox{0.8}{$1$}};
      \node at (0.75,0) {\scalebox{0.8}{$0$}};
      \mpsA{0}{0}{colMPS}
    \end{tikzpicture}=
    \left[\matrix{
      0&1-\vartheta&-1+\vartheta\cr
      \vartheta&0&0\cr
      \vartheta&0&0
    }\right],}\qquad
    \eqalign{
      &\begin{tikzpicture}[baseline={([yshift=-0.6ex]current bounding box.center)},scale=0.5]
        \mpsWire{0}{-0.625}{0}{0.625}
        \gridLine{-0.55}{0}{0.55}{0}
        \node at (-0.75,0) {\scalebox{0.8}{$1$}};
        \node at (0.75,0) {\scalebox{0.8}{$1$}};
        \mpsA{0}{0}{colMPS}
      \end{tikzpicture}=
      \left[\matrix{
        0&1&0\cr
        1&0&0\cr
        0&0&0
      }\right],\\
      &\begin{tikzpicture}[baseline={([yshift=-0.6ex]current bounding box.center)},scale=0.5]
        \mpsWire{0}{-0.625}{0}{0.625}
        \gridLine{-0.55}{0}{0.55}{0}
        \node at (-0.75,0) {\scalebox{0.8}{$s$}};
        \node at (0.75,0) {\scalebox{0.8}{$b$}};
        \mpsB{0}{0}{colMPS}
      \end{tikzpicture}=
      \delta_{b,s}
      \left[\matrix{
        \delta_{s,0} &0&0\cr
        0 &\delta_{s,1}& 0 \cr
        0 & 0 & \delta_{s,1}
      }\right],}
\end{eqnarray}
while the two-step bulk tensor
$\begin{tikzpicture}[baseline={([yshift=-0.6ex]current bounding box.center)},scale=0.5]
  \mpsWire{0}{0.75}{0}{-0.75}
  \gridLine{-0.5}{-0.25}{0.5}{-0.25}
  \gridLine{-0.5}{0.25}{0.5}{0.25}
  \mpsC{0}{-0.25}{0.25}{colMPS}
\end{tikzpicture}$,
and the boundary vectors 
$\begin{tikzpicture}[baseline={([yshift=-0.6ex]current bounding box.center)},scale=0.5]
  \mpsWire{0}{0}{0}{-0.5}
  \mpsBvec{0}{0}
\end{tikzpicture}$,
$\begin{tikzpicture}[baseline={([yshift=-0.6ex]current bounding box.center)},scale=0.5]
  \mpsWire{0}{0}{0}{0.5}
  \mpsBvecW{0}{0}{colMPS}
\end{tikzpicture}$, and
$\begin{tikzpicture}[baseline={([yshift=-0.6ex]current bounding box.center)},scale=0.5]
  \mpsWire{0}{0}{0}{0.5}
  \mpsBvecV{0}{0}{colMPS}
\end{tikzpicture}$, are reported in~\ref{sec:tensorsOther}.
Here the parameter $\vartheta$ is set by the initial state~\eref{eq:defSolvableIS}.

\subsection{Leading eigenvectors of generalized transfer matrices}
We use the ideas summarized above to find leading eigenvectors of
$\mathbb{W}_{\beta,t}^{(\pm)}$. We again assume an MPO representation similar
to~\eref{eq:defAnstazFPmpo},
\begin{eqnarray}\fl
  \sbra{l_{\beta,t}^{(+)}}=
  \begin{tikzpicture}[baseline={([yshift=-0.6ex]current bounding box.center)},scale=0.55]
    \draw[thick,colLines,rounded corners] (0.75,4) -- (-0.75,4) -- (-0.75,6) -- (0.75,6);
    \draw[thick,colLines,rounded corners] (0.75,3) -- (-0.875,3) -- (-0.875,7) -- (0.75,7);
    \draw[thick,colLines,rounded corners] (0.75,2) -- (-1,2) -- (-1,8) -- (0.75,8);
    \draw[thick,colLines,rounded corners] (0.75,1) -- (-1.125,1) -- (-1.125,9) -- (0.75,9);
    \mpsWire{0}{0}{0}{5}
    \vLine{0}{5}{0.75}{5}
    \mpsBvecW{0}{0}{colMPSBeta}
    \mpsA{0}{1}{colMPSBeta}
    \mpsB{0}{2}{colMPSBeta}
    \mpsA{0}{3}{colMPSBeta}
    \mpsB{0}{4}{colMPSBeta}
    \mpsBvecW{0}{5}{colMPSBeta}
    \foreach \x in {0,...,5}{\plus{0}{\x}}
  \end{tikzpicture},\qquad
  \sket{r_{\beta,t}^{(+)}}=
  \begin{tikzpicture}[baseline={([yshift=-0.6ex]current bounding box.center)},scale=0.55]
    \draw[thick,colLines,rounded corners] (-0.75,4) -- (0.75,4) -- (0.75,6) -- (-0.75,6);
    \draw[thick,colLines,rounded corners] (-0.75,3) -- (0.875,3) -- (0.875,7) -- (-0.75,7);
    \draw[thick,colLines,rounded corners] (-0.75,2) -- (1,2) -- (1,8) -- (-0.75,8);
    \draw[thick,colLines,rounded corners] (-0.75,1) -- (1.125,1) -- (1.125,9) -- (-0.75,9);
    \mpsWire{0}{0}{0}{5}
    \vLine{0}{5}{-0.75}{5}
    \mpsBvecV{0}{0}{colMPSBeta}
    \mpsB{0}{1}{colMPSBeta}
    \mpsA{0}{2}{colMPSBeta}
    \mpsB{0}{3}{colMPSBeta}
    \mpsA{0}{4}{colMPSBeta}
    \mpsBvecV{0}{5}{colMPSBeta}
    \foreach \x in {0,...,5}{\plus{0}{\x}}
  \end{tikzpicture},\qquad
  \sbra{l_{\beta,t}^{(-)}}=
  \begin{tikzpicture}[baseline={([yshift=-0.6ex]current bounding box.center)},scale=0.55]
    \draw[thick,colLines,rounded corners] (0.75,4) -- (-0.75,4) -- (-0.75,6) -- (0.75,6);
    \draw[thick,colLines,rounded corners] (0.75,3) -- (-0.875,3) -- (-0.875,7) -- (0.75,7);
    \draw[thick,colLines,rounded corners] (0.75,2) -- (-1,2) -- (-1,8) -- (0.75,8);
    \draw[thick,colLines,rounded corners] (0.75,1) -- (-1.125,1) -- (-1.125,9) -- (0.75,9);
    \mpsWire{0}{0}{0}{5}
    \vLine{0}{5}{0.75}{5}
    \mpsBvecW{0}{0}{colMPSBeta}
    \mpsA{0}{1}{colMPSBeta}
    \mpsB{0}{2}{colMPSBeta}
    \mpsA{0}{3}{colMPSBeta}
    \mpsB{0}{4}{colMPSBeta}
    \mpsBvecW{0}{5}{colMPSBeta}
    \foreach \x in {0,...,5}{\wedgeL{0}{\x}}
  \end{tikzpicture},\qquad
  \sket{r_{\beta,t}^{(-)}}=
  \begin{tikzpicture}[baseline={([yshift=-0.6ex]current bounding box.center)},scale=0.55]
    \draw[thick,colLines,rounded corners] (-0.75,4) -- (0.75,4) -- (0.75,6) -- (-0.75,6);
    \draw[thick,colLines,rounded corners] (-0.75,3) -- (0.875,3) -- (0.875,7) -- (-0.75,7);
    \draw[thick,colLines,rounded corners] (-0.75,2) -- (1,2) -- (1,8) -- (-0.75,8);
    \draw[thick,colLines,rounded corners] (-0.75,1) -- (1.125,1) -- (1.125,9) -- (-0.75,9);
    \mpsWire{0}{0}{0}{5}
    \vLine{0}{5}{-0.75}{5}
    \mpsBvecV{0}{0}{colMPSBeta}
    \mpsB{0}{1}{colMPSBeta}
    \mpsA{0}{2}{colMPSBeta}
    \mpsB{0}{3}{colMPSBeta}
    \mpsA{0}{4}{colMPSBeta}
    \mpsBvecV{0}{5}{colMPSBeta}
    \foreach \x in {0,...,5}{\wedgeR{0}{\x}}
  \end{tikzpicture}.
\end{eqnarray}
Note that in the case of $Q^{(+)}$ we assume the left and right eigenvectors to
be made up of the same bulk tensors, while for $Q^{(-)}$ they change. This is
due to the symmetry of charges: $Q^{(+)}$ is even under left-right reflection,
while $Q^{(-)}$ is odd.\footnote{On the other hand, the top boundary vectors do not share this property: they are generically assumed to be different since the matrices defining the MPO representation of $e^{i\beta Q^{(\pm)}}$ (cf.\ \eref{eq:MPOqmDef} and \eref{eq:MPOqpDef}) are not symmetric.}

To find tensors that constitute fixed points we again assume a set of appropriate local algebraic relations. In particular, the relations fulfilled by the bottom and bulk tensors only (i.e.\ those given by Eq.~\eref{eq:localRelationsLeftBulkBottom}) stay unchanged, while the ones involving top boundary vectors (cf.\ Eq.~\eref{eq:localRelationsLeftTop}) have to be amended to allow for the additional MPO auxiliary space,
\begin{eqnarray}\label{eq:localRelationsLeftBeta}
  \fl
  \begin{tikzpicture}[baseline={([yshift=0.7ex]current bounding box.center)},scale=0.5]
    \mpsWire{0}{0.25}{0}{4}
    \draw[thick,colLines,rounded corners] (1.75,3) -- (-0.5,3) -- (-0.5,5) -- (1.75,5);
    \draw[thick,colLines,rounded corners] (1.75,2) -- (-0.625,2) -- (-0.625,6) -- (1.75,6);
    \draw[thick,colLines,rounded corners] (1.75,1) -- (-0.75,1) -- (-0.75,7) -- (1.75,7);
    \vLine{0}{4}{1.75}{4}
    \mpsBvecW{0}{4}{colMPSBeta}
    \mpsC{0}{1}{2}{colMPSBeta}
    \mpsB{0}{3}{colMPSBeta}
    \gridLine{1}{2}{1}{6}
    \sCircle{1}{2}{colU}
    \bCircle{1}{3}{colU}
    \vmpsNS{1}{4}{colObs}{colLines}
    \bCircle{1}{5}{colUc}
    \sCircle{1}{6}{colUc}
    \foreach \x in {1,...,4}{\plus{0}{\x}}
  \end{tikzpicture}=
  \begin{tikzpicture}[baseline={([yshift=0.7ex]current bounding box.center)},scale=0.5]
    \mpsWire{0}{0.25}{0}{4}
    \draw[thick,colLines,rounded corners] (0.75,3) -- (-0.5,3) -- (-0.5,5) -- (0.75,5);
    \draw[thick,colLines,rounded corners] (0.75,2) -- (-0.625,2) -- (-0.625,6) -- (0.75,6);
    \draw[thick,colLines,rounded corners] (0.75,1) -- (-0.75,1) -- (-0.75,7) -- (0.75,7);
    \vLine{0}{4}{0.75}{4}
    \mpsBvecV{0}{4}{colMPSBeta}
    \mpsA{0}{1}{colMPSBeta}
    \mpsB{0}{2}{colMPSBeta}
    \mpsA{0}{3}{colMPSBeta}
    \foreach \x in {1,...,4}{\plus{0}{\x}}
  \end{tikzpicture},\quad
  \begin{tikzpicture}[baseline={([yshift=0.7ex]current bounding box.center)},scale=0.5]
    \mpsWire{0}{0.25}{0}{3}
    \draw[thick,colLines,rounded corners] (1.75,2) -- (-0.5,2) -- (-0.5,4) -- (1.75,4);
    \draw[thick,colLines,rounded corners] (1.75,1) -- (-0.625,1) -- (-0.625,5) -- (1.75,5);
    \vLine{0}{3}{1.75}{3}
    \mpsBvecV{0}{3}{colMPSBeta}
    \mpsC{0}{1}{2}{colMPSBeta}
    \gridLine{1}{2}{1}{4}
    \sCircle{1}{2}{colU}
    \vmpsNS{1}{3}{colObs}{colLines}
    \sCircle{1}{4}{colUc}
    \foreach \x in {1,...,3}{\plus{0}{\x}}
  \end{tikzpicture}=
  \Lambda_{\beta}^{(+)}
  \begin{tikzpicture}[baseline={([yshift=0.7ex]current bounding box.center)},scale=0.5]
    \mpsWire{0}{0.25}{0}{3}
    \draw[thick,colLines,rounded corners] (0.75,2) -- (-0.5,2) -- (-0.5,4) -- (0.75,4);
    \draw[thick,colLines,rounded corners] (0.75,1) -- (-0.625,1) -- (-0.625,5) -- (0.75,5);
    \vLine{0}{3}{0.75}{3}
    \mpsBvecW{0}{3}{colMPSBeta}
    \mpsB{0}{2}{colMPSBeta}
    \mpsA{0}{1}{colMPSBeta}
    \foreach \x in {1,...,3}{\plus{0}{\x}}
  \end{tikzpicture},\qquad
  \begin{tikzpicture}[baseline={([yshift=0.7ex]current bounding box.center)},scale=0.5]
    \mpsWire{0}{0.25}{0}{4}
    \draw[thick,colLines,rounded corners] (1.75,3) -- (-0.5,3) -- (-0.5,5) -- (1.75,5);
    \draw[thick,colLines,rounded corners] (1.75,2) -- (-0.625,2) -- (-0.625,6) -- (1.75,6);
    \draw[thick,colLines,rounded corners] (1.75,1) -- (-0.75,1) -- (-0.75,7) -- (1.75,7);
    \vLine{0}{4}{1.75}{4}
    \mpsBvecW{0}{4}{colMPSBeta}
    \mpsC{0}{1}{2}{colMPSBeta}
    \mpsB{0}{3}{colMPSBeta}
    \gridLine{1}{2}{1}{6}
    \sCircle{1}{2}{colU}
    \bCircle{1}{3}{colU}
    \vmpsV{1}{4}{colObs}{colLines}
    \bCircle{1}{5}{colUc}
    \sCircle{1}{6}{colUc}
    \foreach \x in {1,...,4}{\wedgeL{0}{\x}}
  \end{tikzpicture}=
  \begin{tikzpicture}[baseline={([yshift=0.7ex]current bounding box.center)},scale=0.5]
    \mpsWire{0}{0.25}{0}{4}
    \draw[thick,colLines,rounded corners] (0.75,3) -- (-0.5,3) -- (-0.5,5) -- (0.75,5);
    \draw[thick,colLines,rounded corners] (0.75,2) -- (-0.625,2) -- (-0.625,6) -- (0.75,6);
    \draw[thick,colLines,rounded corners] (0.75,1) -- (-0.75,1) -- (-0.75,7) -- (0.75,7);
    \vLine{0}{4}{0.75}{4}
    \mpsBvecV{0}{4}{colMPSBeta}
    \mpsA{0}{1}{colMPSBeta}
    \mpsB{0}{2}{colMPSBeta}
    \mpsA{0}{3}{colMPSBeta}
    \foreach \x in {1,...,4}{\wedgeL{0}{\x}}
  \end{tikzpicture},\quad
  \begin{tikzpicture}[baseline={([yshift=0.7ex]current bounding box.center)},scale=0.5]
    \mpsWire{0}{0.25}{0}{3}
    \draw[thick,colLines,rounded corners] (1.75,2) -- (-0.5,2) -- (-0.5,4) -- (1.75,4);
    \draw[thick,colLines,rounded corners] (1.75,1) -- (-0.625,1) -- (-0.625,5) -- (1.75,5);
    \vLine{0}{3}{1.75}{3}
    \mpsBvecV{0}{3}{colMPSBeta}
    \mpsC{0}{1}{2}{colMPSBeta}
    \gridLine{1}{2}{1}{4}
    \sCircle{1}{2}{colU}
    \vmpsW{1}{3}{colObs}{colLines}
    \sCircle{1}{4}{colUc}
    \foreach \x in {1,...,3}{\wedgeL{0}{\x}}
  \end{tikzpicture}=
  \Lambda_{\beta}^{(-)}
  \begin{tikzpicture}[baseline={([yshift=0.7ex]current bounding box.center)},scale=0.5]
    \mpsWire{0}{0.25}{0}{3}
    \draw[thick,colLines,rounded corners] (0.75,2) -- (-0.5,2) -- (-0.5,4) -- (0.75,4);
    \draw[thick,colLines,rounded corners] (0.75,1) -- (-0.625,1) -- (-0.625,5) -- (0.75,5);
    \vLine{0}{3}{0.75}{3}
    \mpsBvecW{0}{3}{colMPSBeta}
    \mpsB{0}{2}{colMPSBeta}
    \mpsA{0}{1}{colMPSBeta}
    \foreach \x in {1,...,3}{\wedgeL{0}{\x}}
  \end{tikzpicture},
\end{eqnarray}
while the corresponding set of relations for the top boundary vectors of \emph{right} fixed points read as
\begin{eqnarray}\label{eq:localRelationsRightBeta}\fl
  \begin{tikzpicture}[baseline={([yshift=0.7ex]current bounding box.center)},scale=0.5]
    \mpsWire{0}{0.25}{0}{4}
    \draw[thick,colLines,rounded corners] (-1.75,3) -- (0.5,3) -- (0.5,5) -- (-1.75,5);
    \draw[thick,colLines,rounded corners] (-1.75,2) -- (0.625,2) -- (0.625,6) -- (-1.75,6);
    \draw[thick,colLines,rounded corners] (-1.75,1) -- (0.75,1) -- (0.75,7) -- (-1.75,7);
    \vLine{0}{4}{-1.75}{4}
    \mpsBvecW{0}{4}{colMPSBeta}
    \mpsC{0}{1}{2}{colMPSBeta}
    \mpsB{0}{3}{colMPSBeta}
    \gridLine{-1}{2}{-1}{6}
    \sCircle{-1}{2}{colU}
    \bCircle{-1}{3}{colU}
    \vmpsNS{-1}{4}{colObs}{colLines}
    \bCircle{-1}{5}{colUc}
    \sCircle{-1}{6}{colUc}
    \foreach \x in {1,...,4}{\plus{0}{\x}}
  \end{tikzpicture}=
  \begin{tikzpicture}[baseline={([yshift=0.7ex]current bounding box.center)},scale=0.5]
    \mpsWire{0}{0.25}{0}{4}
    \draw[thick,colLines,rounded corners] (-0.75,3) -- (0.5,3) -- (0.5,5) -- (-0.75,5);
    \draw[thick,colLines,rounded corners] (-0.75,2) -- (0.625,2) -- (0.625,6) -- (-0.75,6);
    \draw[thick,colLines,rounded corners] (-0.75,1) -- (0.75,1) -- (0.75,7) -- (-0.75,7);
    \vLine{0}{4}{-0.75}{4}
    \mpsBvecV{0}{4}{colMPSBeta}
    \mpsA{0}{1}{colMPSBeta}
    \mpsB{0}{2}{colMPSBeta}
    \mpsA{0}{3}{colMPSBeta}
    \foreach \x in {1,...,4}{\plus{0}{\x}}
  \end{tikzpicture},\quad
  \begin{tikzpicture}[baseline={([yshift=0.7ex]current bounding box.center)},scale=0.5]
    \mpsWire{0}{0.25}{0}{3}
    \draw[thick,colLines,rounded corners] (-1.75,2) -- (0.5,2) -- (0.5,4) -- (-1.75,4);
    \draw[thick,colLines,rounded corners] (-1.75,1) -- (0.625,1) -- (0.625,5) -- (-1.75,5);
    \vLine{0}{3}{-1.75}{3}
    \mpsBvecV{0}{3}{colMPSBeta}
    \mpsC{0}{1}{2}{colMPSBeta}
    \gridLine{-1}{2}{-1}{4}
    \sCircle{-1}{2}{colU}
    \vmpsNS{-1}{3}{colObs}{colLines}
    \sCircle{-1}{4}{colUc}
    \foreach \x in {1,...,3}{\plus{0}{\x}}
  \end{tikzpicture}=
  \Lambda_{\beta}^{(+)}
  \begin{tikzpicture}[baseline={([yshift=0.7ex]current bounding box.center)},scale=0.5]
    \mpsWire{0}{0.25}{0}{3}
    \draw[thick,colLines,rounded corners] (-0.75,2) -- (0.5,2) -- (0.5,4) -- (-0.75,4);
    \draw[thick,colLines,rounded corners] (-0.75,1) -- (0.625,1) -- (0.625,5) -- (-0.75,5);
    \vLine{0}{3}{-0.75}{3}
    \mpsBvecW{0}{3}{colMPSBeta}
    \mpsB{0}{2}{colMPSBeta}
    \mpsA{0}{1}{colMPSBeta}
    \foreach \x in {1,...,3}{\plus{0}{\x}}
  \end{tikzpicture},\qquad
  \begin{tikzpicture}[baseline={([yshift=0.7ex]current bounding box.center)},scale=0.5]
    \mpsWire{0}{0.25}{0}{4}
    \draw[thick,colLines,rounded corners] (-1.75,3) -- (0.5,3) -- (0.5,5) -- (-1.75,5);
    \draw[thick,colLines,rounded corners] (-1.75,2) -- (0.625,2) -- (0.625,6) -- (-1.75,6);
    \draw[thick,colLines,rounded corners] (-1.75,1) -- (0.75,1) -- (0.75,7) -- (-1.75,7);
    \vLine{0}{4}{-1.75}{4}
    \mpsBvecW{0}{4}{colMPSBeta}
    \mpsC{0}{1}{2}{colMPSBeta}
    \mpsB{0}{3}{colMPSBeta}
    \gridLine{-1}{2}{-1}{6}
    \sCircle{-1}{2}{colU}
    \bCircle{-1}{3}{colU}
    \vmpsV{-1}{4}{colObs}{colLines}
    \bCircle{-1}{5}{colUc}
    \sCircle{-1}{6}{colUc}
    \foreach \x in {1,...,4}{\wedgeR{0}{\x}}
  \end{tikzpicture}=
  \begin{tikzpicture}[baseline={([yshift=0.7ex]current bounding box.center)},scale=0.5]
    \mpsWire{0}{0.25}{0}{4}
    \draw[thick,colLines,rounded corners] (-0.75,3) -- (0.5,3) -- (0.5,5) -- (-0.75,5);
    \draw[thick,colLines,rounded corners] (-0.75,2) -- (0.625,2) -- (0.625,6) -- (-0.75,6);
    \draw[thick,colLines,rounded corners] (-0.75,1) -- (0.75,1) -- (0.75,7) -- (-0.75,7);
    \vLine{0}{4}{-0.75}{4}
    \mpsBvecV{0}{4}{colMPSBeta}
    \mpsA{0}{1}{colMPSBeta}
    \mpsB{0}{2}{colMPSBeta}
    \mpsA{0}{3}{colMPSBeta}
    \foreach \x in {1,...,4}{\wedgeR{0}{\x}}
  \end{tikzpicture},\quad
  \begin{tikzpicture}[baseline={([yshift=0.7ex]current bounding box.center)},scale=0.5]
    \mpsWire{0}{0.25}{0}{3}
    \draw[thick,colLines,rounded corners] (-1.75,2) -- (0.5,2) -- (0.5,4) -- (-1.75,4);
    \draw[thick,colLines,rounded corners] (-1.75,1) -- (0.625,1) -- (0.625,5) -- (-1.75,5);
    \vLine{0}{3}{-1.75}{3}
    \mpsBvecV{0}{3}{colMPSBeta}
    \mpsC{0}{1}{2}{colMPSBeta}
    \gridLine{-1}{2}{-1}{4}
    \sCircle{-1}{2}{colU}
    \vmpsW{-1}{3}{colObs}{colLines}
    \sCircle{-1}{4}{colUc}
    \foreach \x in {1,...,3}{\wedgeR{0}{\x}}
  \end{tikzpicture}=
  \Lambda_{\beta}^{(-)}
  \begin{tikzpicture}[baseline={([yshift=0.7ex]current bounding box.center)},scale=0.5]
    \mpsWire{0}{0.25}{0}{3}
    \draw[thick,colLines,rounded corners] (-0.75,2) -- (0.5,2) -- (0.5,4) -- (-0.75,4);
    \draw[thick,colLines,rounded corners] (-0.75,1) -- (0.625,1) -- (0.625,5) -- (-0.75,5);
    \vLine{0}{3}{-0.75}{3}
    \mpsBvecW{0}{3}{colMPSBeta}
    \mpsB{0}{2}{colMPSBeta}
    \mpsA{0}{1}{colMPSBeta}
    \foreach \x in {1,...,3}{\wedgeR{0}{\x}}
  \end{tikzpicture}.
\end{eqnarray}
Combining the relevant equations we obtain a set of algebraic relations
involving a finite number of variables. Upon solving these we again find
a solution with a $3$-dimensional representation, where the bulk tensors
are simply related to the original ones~\eref{eq:originalFixedPointTensors},
\begin{eqnarray}\label{eq:newFixedPointTensors}
  \begin{tikzpicture}[baseline={([yshift=-0.6ex]current bounding box.center)},scale=0.5]
    \mpsWire{0}{-0.625}{0}{0.625}
    \gridLine{-0.55}{0}{0.55}{0}
    \mpsA{0}{0}{colMPSBeta}
    \plus{0}{0}
  \end{tikzpicture}=
  \begin{tikzpicture}[baseline={([yshift=-0.6ex]current bounding box.center)},scale=0.5]
    \mpsWire{0}{-0.625}{0}{0.625}
    \gridLine{-0.55}{0}{0.55}{0}
    \mpsA{0}{0}{colMPSBeta}
    \wedgeL{0}{0}
  \end{tikzpicture}=
  \begin{tikzpicture}[baseline={([yshift=-0.6ex]current bounding box.center)},scale=0.5]
    \mpsWire{0}{-0.625}{0}{0.625}
    \gridLine{-0.55}{0}{0.55}{0}
    \mpsA{0}{0}{colMPSBeta}
    \wedgeR{0}{0}
  \end{tikzpicture}=
  \begin{tikzpicture}[baseline={([yshift=-0.6ex]current bounding box.center)},scale=0.5]
    \mpsWire{0}{-0.625}{0}{0.625}
    \gridLine{-0.55}{0}{0.55}{0}
    \mpsA{0}{0}{colMPS}
  \end{tikzpicture},\quad
  \begin{tikzpicture}[baseline={([yshift=-0.6ex]current bounding box.center)},scale=0.5]
    \mpsWire{0}{-0.625}{0}{0.625}
    \gridLine{-0.55}{0}{0.55}{0}
    \mpsB{0}{0}{colMPSBeta}
    \plus{0}{0}
  \end{tikzpicture}=
  \begin{tikzpicture}[baseline={([yshift=-1.2ex]current bounding box.center)},scale=0.5]
    \mpsWire{0}{-0.625}{0}{1}
    \gridLine{-0.55}{0}{0.55}{0}
    \mpsB{0}{0}{colMPS}
    \mpsAux{0}{0.625}{colMPSBeta}
    \plus{0}{0.625}
  \end{tikzpicture},\quad
  \begin{tikzpicture}[baseline={([yshift=-0.6ex]current bounding box.center)},scale=0.5]
    \mpsWire{0}{-0.625}{0}{0.625}
    \gridLine{-0.55}{0}{0.55}{0}
    \mpsB{0}{0}{colMPSBeta}
    \wedgeL{0}{0}
  \end{tikzpicture}=
  \begin{tikzpicture}[baseline={([yshift=-1.2ex]current bounding box.center)},scale=0.5]
    \mpsWire{0}{-0.625}{0}{1}
    \gridLine{-0.55}{0}{0.55}{0}
    \mpsB{0}{0}{colMPS}
    \mpsAux{0}{0.625}{colMPSBeta}
    \wedgeL{0}{0.625}
  \end{tikzpicture},\quad
  \begin{tikzpicture}[baseline={([yshift=-0.6ex]current bounding box.center)},scale=0.5]
    \mpsWire{0}{-0.625}{0}{0.625}
    \gridLine{-0.55}{0}{0.55}{0}
    \mpsB{0}{0}{colMPSBeta}
    \wedgeR{0}{0}
  \end{tikzpicture}=
  \begin{tikzpicture}[baseline={([yshift=-1.2ex]current bounding box.center)},scale=0.5]
    \mpsWire{0}{-0.625}{0}{1}
    \gridLine{-0.55}{0}{0.55}{0}
    \mpsB{0}{0}{colMPS}
    \mpsAux{0}{0.625}{colMPSBeta}
    \wedgeR{0}{0.625}
  \end{tikzpicture},
\end{eqnarray}
where
$\begin{tikzpicture}[baseline={([yshift=-0.6ex]current bounding box.center)},scale=0.5]
  \mpsWire{0}{0.1}{0}{0.9}
  \mpsAux{0}{0.5}{colMPSBeta}
  \wedgeL{0}{0.5}
\end{tikzpicture}$,
$\begin{tikzpicture}[baseline={([yshift=-0.6ex]current bounding box.center)},scale=0.5]
  \mpsWire{0}{0.1}{0}{0.9}
  \mpsAux{0}{0.5}{colMPSBeta}
  \wedgeR{0}{0.5}
\end{tikzpicture}$, and
$\begin{tikzpicture}[baseline={([yshift=-0.6ex]current bounding box.center)},scale=0.5]
  \mpsWire{0}{0.1}{0}{0.9}
  \mpsAux{0}{0.5}{colMPSBeta}
  \plus{0}{0.5}
\end{tikzpicture}$, are diagonal matrices in the auxiliary space,
\begin{eqnarray}\label{eq:defPentagon}\fl
  \begin{tikzpicture}[baseline={([yshift=-0.6ex]current bounding box.center)},scale=0.5]
    \mpsWire{0}{-0.5}{0}{0.5}
    \mpsAux{0}{0}{colMPSBeta}
  \end{tikzpicture}=
  \left[\matrix{
    1 & 0 & 0 \cr
    0 & \gamma & 0 \cr
    0 & 0 & \gamma
    }\right],\qquad
    \begin{tikzpicture}[baseline={([yshift=-0.6ex]current bounding box.center)},scale=0.5]
      \mpsWire{0}{-0.5}{0}{0.5}
      \mpsAux{0}{0}{colMPSBeta}
      \wedgeL{0}{0}
    \end{tikzpicture}=
    \left.
    \begin{tikzpicture}[baseline={([yshift=-0.6ex]current bounding box.center)},scale=0.5]
      \mpsWire{0}{-0.5}{0}{0.5}
      \mpsAux{0}{0}{colMPSBeta}
    \end{tikzpicture}\right|_{\gamma\to e^{-i\beta}},\qquad
    \begin{tikzpicture}[baseline={([yshift=-0.6ex]current bounding box.center)},scale=0.5]
      \mpsWire{0}{-0.5}{0}{0.5}
      \mpsAux{0}{0}{colMPSBeta}
      \wedgeR{0}{0}
    \end{tikzpicture}=
    \left.
    \begin{tikzpicture}[baseline={([yshift=-0.6ex]current bounding box.center)},scale=0.5]
      \mpsWire{0}{-0.5}{0}{0.5}
      \mpsAux{0}{0}{colMPSBeta}
    \end{tikzpicture}\right|_{\gamma\to e^{i\beta}},\qquad
    \begin{tikzpicture}[baseline={([yshift=-0.6ex]current bounding box.center)},scale=0.5]
      \mpsWire{0}{-0.5}{0}{0.5}
      \mpsAux{0}{0}{colMPSBeta}
      \plus{0}{0}
    \end{tikzpicture}=
    \left.
    \begin{tikzpicture}[baseline={([yshift=-0.6ex]current bounding box.center)},scale=0.5]
      \mpsWire{0}{-0.5}{0}{0.5}
      \mpsAux{0}{0}{colMPSBeta}
    \end{tikzpicture}\right|_{\gamma\to e^{i\beta}\Lambda_{\beta}^{(+)}}.
\end{eqnarray}
Note that the bulk tensors in $\sbra{l_{\beta,t}^{(-)}}$ and 
$\sket{r_{\beta,t}^{(-)}}$ are related to each other through a flip of sign of $\beta$.
The remaining tensors and boundary vectors are reported in~\ref{sec:tensorsOther}.
The corresponding eigenvalues are determined as (see \ref{sec:uniquenessFPs}),
\begin{eqnarray}
  \log \Lambda_{\beta}^{(\pm)}=
  \lim_{L\to\infty}
  \frac{1}{L}\log \mel{\Psi_0}{e^{i\beta Q^{(\pm)}}}{\Psi_0},
\end{eqnarray}
and are
\begin{eqnarray}\label{eq:defLambda}
  \Lambda_{\beta}^{(+)}=1-\vartheta+\vartheta e^{2i\beta},\qquad
  \Lambda_{\beta}^{(-)}=1.
\end{eqnarray}

Finally, we note that the fixed points are for convenience \emph{not} normalized, but rather
we keep the overlap between the right and left eigenvectors to be 
\begin{eqnarray}\label{eq:FPnorm}
  \sbraket{l_{\beta,t}^{(+)}}{r_{\beta,t}^{(+)}}=\left.\Lambda_{\beta}^{(+)}\right.^{2t},
  \qquad
  \sbraket{l_{\beta,t}^{(-)}}{r_{\beta,t}^{(-)}}=1=\left.\Lambda_{\beta}^{(-)}\right.^{2t}.
\end{eqnarray}
See~\ref{sec:FPnormalization} for details.

\section{Asymptotics of charged moments}\label{sec:asymptotics}
When the subsystem size $l$ and the system size $L$ are sufficiently large
compared to the time $t$ (in particular, when $l,L>2t+1$) the charged moments
factorize into the two contributions corresponding to the edges of the
subsystem,
\begin{eqnarray}
  \lim_{l,L\to\infty} \frac{Z_{\alpha,\vec{\beta}}^{(\pm)}(l,t)}
  {\prod_{j=1}^{\alpha}
  \left.\Lambda_{\beta_{j}}^{(\pm)}\right.^{l}}
  =
  \frac{z_{\alpha,\vec{\beta}}^{{\rm L}\,(\pm)}(t)\,
  z_{\alpha,\vec{\beta}}^{{\rm R}\,(\pm)}(t)}
  {\prod_{j=1}^{\alpha}
  \left.\Lambda_{\beta_{j}}^{(\pm)}\right.^{2t}},
\end{eqnarray}
where $z_{\alpha,\vec{\beta}}^{{\rm L/R}\, (\pm)}$ are the contributions
given by the (un-normalized) leading eigenvectors,
\begin{eqnarray}\fl
  \eqalign{
    z_{\alpha,\vec{\beta}}^{{\rm L}\, (\pm)}=
    \sbra{l_{t}}^{\otimes \alpha}\,
    \mathbb{P}_{\alpha,t}
    \left(
    \mathbb{B}_{{\rm L},\beta_1,t}^{(\pm)}\sket{r_{\beta_1,t}^{(\pm)}}
    \otimes
    \mathbb{B}_{{\rm L},\beta_2,t}^{(\pm)}\sket{r_{\beta_2,t}^{(\pm)}}
    \otimes
    \cdots
    \otimes
    \mathbb{B}_{{\rm L},\beta_{\alpha},t}^{(\pm)}\sket{r_{\beta_{\alpha},t}^{(\pm)}}\right),\\
    z_{\alpha,\vec{\beta}}^{{\rm R}\, (\pm)}=
    \left(
    \sbra{l_{\beta_1,t}^{(\pm)}}\mathbb{B}_{{\rm R},\beta_1,t}^{(\pm)}
    \otimes
    \sbra{l_{\beta_2,t}^{(\pm)}}\mathbb{B}_{{\rm R},\beta_2,t}^{(\pm)}
    \otimes
    \cdots
    \otimes
    \sbra{l_{\beta_{\alpha},t}^{(\pm)}}\mathbb{B}_{{\rm R},\beta_{\alpha},t}^{(\pm)}
    \right)\,
    \mathbb{P}_{\alpha,t}^{\dagger}
    \sket{r_{t}}^{\otimes \alpha}.
    }
\end{eqnarray}
In terms of the fixed-point tensors, these expressions take form of
$2t\times 2\alpha$ tensor networks,
\begin{eqnarray} \label{eq:defTNOneEdgeContrib}
  z_{\alpha,\vec{\beta}}^{{\rm L}\, (-)}(t)
  =
  \begin{tikzpicture}[baseline={([yshift=-0.6ex]current bounding box.center)},scale=0.5]
    \foreach \t in {1,...,4}{
      \gridLine{0.4}{\t}{6.6}{\t}
      \gridLine{0.4}{\t-0.4}{6.6}{\t-0.4}
      \leftHook{0.4}{\t}
      \rightHook{6.6}{\t}
    }
    \foreach \x in {1,...,6}{\mpsWire{\x}{0}{\x}{5}}
    \foreach \x in {2,4,6}{
      \foreach \t in {1,3}{
        \mpsAuxWhite{\x}{\t+0.45}
      }
    }
    \foreach \x in {1,3,5}{
      \mpsBvecW{\x}{0}{colMPS}
      \mpsBvec{\x}{5}
      \mpsBvecV{\x+1}{0}{colMPS}
      \foreach \t in {1,3}{
        \mpsA{\x}{\t}{colMPS}
        \mpsB{\x+1}{\t}{colMPS}
        \mpsB{\x}{\t+1}{colMPS}
        \mpsA{\x+1}{\t+1}{colMPS}
      }
    }
    \foreach \x in {2,4,6}{
      \vLine{\x}{5}{\x-0.5}{5}
      \vLine{\x-0.5}{5.25}{\x-0.5}{4.75}
      \mpsBvecV{\x}{5}{colMPSBeta}
      \wedgeR{\x}{5}
      \foreach \t in {1,3}{
        \mpsAux{\x}{\t+0.45}{colMPSBeta}
        \wedgeR{\x}{\t+0.45}
      }
    }
    \node at (2,-0.75) {\scalebox{0.8}{$\beta_1$}};
    \node at (4,-0.75) {\scalebox{0.8}{$\cdots$}};
    \node at (6,-0.75) {\scalebox{0.8}{$\beta_{\alpha}$}};
  \end{tikzpicture},\qquad
  z_{\alpha,\vec{\beta}}^{{\rm R}\, (-)}(t)
  =
  \begin{tikzpicture}[baseline={([yshift=-0.6ex]current bounding box.center)},scale=0.5]
    \foreach \t in {1,...,4}{
      \gridLine{0.4}{\t}{6.6}{\t}
      \gridLine{0.4}{\t-0.4}{6.6}{\t-0.4}
      \leftHook{0.4}{\t}
      \rightHook{6.6}{\t}
    }
    \foreach \x in {1,...,6}{\mpsWire{\x}{0}{\x}{5}}
    \foreach \x in {1,3,5}{
      \foreach \t in {2}{
        \mpsAuxWhite{\x}{\t+0.45}
      }
    }
    \foreach \x in {1,3,5}{
      \mpsBvecW{\x}{0}{colMPS}
      \mpsBvec{\x+1}{5}
      \mpsBvecV{\x+1}{0}{colMPS}
      \foreach \t in {1,3}{
        \mpsA{\x}{\t}{colMPS}
        \mpsB{\x+1}{\t}{colMPS}
        \mpsB{\x}{\t+1}{colMPS}
        \mpsA{\x+1}{\t+1}{colMPS}
      }
    }
    \foreach \x in {1,3,5}{
      \vLine{\x}{5}{\x+0.5}{5}
      \vLine{\x+0.5}{5.25}{\x+0.5}{4.75}
      \mpsBvecW{\x}{5}{colMPSBeta}
      \wedgeL{\x}{5}
      \foreach \t in {2}{
        \mpsAux{\x}{\t+0.45}{colMPSBeta}
        \wedgeL{\x}{\t+0.45}
      }
      \foreach \t in {4}{
        \mpsAux{\x}{\t+0.5}{colMPSBeta}
        \wedgeL{\x}{\t+0.5}
      }
    }
    \node at (1,-0.75) {\scalebox{0.8}{$\beta_1$}};
    \node at (3,-0.75) {\scalebox{0.8}{$\cdots$}};
    \node at (5,-0.75) {\scalebox{0.8}{$\beta_{\alpha}$}};
  \end{tikzpicture},
\end{eqnarray}
where the label under a column denotes the value of $\beta$ in which the
tensors in that column are evaluated. The diagrams representing
$z_{\alpha,\vec{\beta}}^{{\rm L/R} (+)}(t)$ are analogous, only the tensors
with $\wLt$ and $\wRt$ are replaced with the corresponding $\wPlust$ tensors.

These tensor networks are hard to contract for finite $t$, but we are able to
characterize their large-$t$ asymptotics. We start by introducing the transfer
matrix $\mathcal{T}_{\alpha,\vec{\gamma}}$, defined as
\begin{eqnarray}
  \mathcal{T}_{\alpha,\vec{\gamma}}:=
  \begin{tikzpicture}[baseline={([yshift=0ex]current bounding box.center)},scale=0.5]
    \foreach \t in {1,2}{
      \gridLine{0.4}{\t}{6.6}{\t}
      \gridLine{0.4}{\t-0.4}{6.6}{\t-0.4}
      \leftHook{0.4}{\t}
      \rightHook{6.6}{\t}
    }
    \foreach \x in {1,...,6}{\mpsWire{\x}{0.25}{\x}{2.75}}
    \foreach \x in {2,4,6}{\mpsAuxWhite{\x}{1.45}}
    \foreach \x in {1,3,5}{
      \foreach \t in {1}{
        \mpsA{\x}{\t}{colMPS}
        \mpsB{\x+1}{\t}{colMPS}
        \mpsB{\x}{\t+1}{colMPS}
        \mpsA{\x+1}{\t+1}{colMPS}
      }
    }
    \foreach \x in {2,4,6}{\mpsAux{\x}{1.45}{colMPSBeta}}
    \node at (2,0) {\scalebox{0.8}{$\gamma_1$}};
    \node at (4,0) {\scalebox{0.8}{$\cdots$}};
    \node at (6,0) {\scalebox{0.8}{$\gamma_{\alpha}$}};
  \end{tikzpicture},
\end{eqnarray}
with generic values of $\gamma_{j}$.  All four relevant tensor networks can be
reproduced by repeatedly applying $\mathcal{T}_{\alpha,\vec{\gamma}}$ and
appropriately setting bottom and top vectors. Moreover, assuming that
$\mathcal{T}_{\alpha,\vec{\gamma}}$ has a unique leading eigenvalue
$\lambda_{\alpha,\vec{\gamma}}$, the boundary vectors play no role in the
asymptotic behaviour of left and right contributions. In this case the
asymptotic slopes of the logarithms of $z^{{\rm L/R}
(\pm)}_{\alpha,\vec{\beta}}$ are completely determined by the leading
eigenvalue of the transfer matrix,
\begin{eqnarray}
  s_{\alpha,\vec{\beta}}^{(+)}:=
  \lim_{t\to\infty} \frac{\log z^{{\rm R} (+)}_{\alpha,\vec{\beta}}(t)}{t}=
  \lim_{t\to\infty} \frac{\log z^{{\rm L} (+)}_{\alpha,\vec{\beta}}(t)}{t}=
  \left.
  \log \lambda_{\alpha,\vec{\gamma}}
  \right|_{\gamma_j\to e^{i\beta_j} \Lambda_{\beta_j}^{(+)}},
\end{eqnarray}
and similarly
\begin{eqnarray}
  s_{\alpha,\vec{\beta}}^{(-)}:=
  \lim_{t\to\infty} \frac{\log z^{{\rm R} (-)}_{\alpha,\vec{\beta}}(t)}{t}=
  \lim_{t\to\infty} \frac{\log z^{{\rm L} (-)}_{\alpha,-\vec{\beta}}(t)}{t}=
  \left.
  \log \lambda_{\alpha,\vec{\gamma}}
  \right|_{\gamma_j\to e^{-i\beta_j}}.
\end{eqnarray}
Note that the symmetry of $Q^{(-)}$ again implies the flip of the sign of
$\vec{\beta}$ between the contributions of the left and right boundary.

\subsection{Spectrum of the transfer matrix}
The spectrum of $\mathcal{T}_{\alpha,\vec{\gamma}}$ can be found by following
and appropriately modifying the approach of Ref.~\cite{klobas2021entanglement}. We first 
clump together two vertical legs and two horizontal legs into single degrees of freedom,
\begin{eqnarray}
  \begin{tikzpicture}[baseline={([yshift=-0.6ex]current bounding box.center)},scale=0.5]
    \foreach \t in {1,2}{\gridLine{0.4}{\t}{2.6}{\t}}
    \foreach \x in {1,2}{\mpsWire{\x}{0.25}{\x}{2.75}}
    \foreach \x in {1}{
      \foreach \t in {1}{
        \mpsA{\x}{\t}{colMPS}
        \mpsB{\x+1}{\t}{colMPS}
        \mpsB{\x}{\t+1}{colMPS}
        \mpsA{\x+1}{\t+1}{colMPS}
      }
    }
    \mpsAux{2}{1.45}{colMPSBeta}
    \node at (0,2) {\scalebox{0.8}{$s_1$}};
    \node at (0,1) {\scalebox{0.8}{$s_2$}};
    \node at (3,2) {\scalebox{0.8}{$b_1$}};
    \node at (3,1) {\scalebox{0.8}{$b_2$}};
    \node at (1,0) {\scalebox{0.8}{$v_1$}};
    \node at (2,0) {\scalebox{0.8}{$v_2$}};
    \node at (1,3) {\scalebox{0.8}{$w_1$}};
    \node at (2,3) {\scalebox{0.8}{$w_2$}};
  \end{tikzpicture}=:
  \begin{tikzpicture}[baseline={([yshift=-0.6ex]current bounding box.center)},scale=0.5]
    \vLine{-0.8}{0}{0.8}{0}
    \vLine{0}{-0.8}{0}{0.8}
    \tenM{0}{0}{colM}
    \node at (0,1) {\scalebox{0.8}{$3 w_1\!+\!w_2$}};
    \node at (0,-1) {\scalebox{0.8}{$3 v_1\!+\!v_2$}};
    \node at (-1,0) {\scalebox{0.8}{$2 s_1\!+\!s_2\mkern48mu$}};
    \node at (1,0) {\scalebox{0.8}{$\mkern48mu 2 b_1\!+\!b_2$}};
  \end{tikzpicture},
\end{eqnarray}
so that the transfer matrix is expressed as~\footnote{Note that we are
suppressing the subscripts $\gamma_j$ to lighten the notation. Everywhere in
this section each column is assumed to correspond to a different value of $\gamma_j$
unless explicitly stated.}
\begin{eqnarray}
  \mathcal{T}_{\alpha,\vec{\gamma}}=
  \begin{tikzpicture}[baseline={([yshift=-0.6ex]current bounding box.center)},scale=0.5]
    \vLine{-0.5}{-0.5}{2.5}{-0.5}
    \vLine{-0.5}{0}{2.5}{0}
    \vleftHook{-0.5}{0}
    \vrightHook{2.5}{0}
    \foreach \x in {0,...,2}
    {
      \fill[white] (\x-0.2,-1) rectangle (\x+0.2,1);
      \vLine{\x}{-1}{\x}{1}
      \tenM{\x}{0}{colM}
    }
  \end{tikzpicture}.
\end{eqnarray}
We then perform a convenient local similarity transformation,
\begin{eqnarray}\label{eq:transformedTensorsENT}
  \begin{tikzpicture}[baseline={([yshift=-0.6ex]current bounding box.center)},scale=0.5]
    \vLine{-0.8}{0}{0.8}{0}
    \vLine{0}{-0.8}{0}{0.8}
    \tenM{0}{0}{colMt}
  \end{tikzpicture}
  =
  \begin{tikzpicture}[baseline={([yshift=-0.6ex]current bounding box.center)},scale=0.5]
    \vLine{-0.8}{0}{0.8}{0}
    \vLine{0}{-1.75}{0}{1.75}
    \draw[colObsLines,fill=white] (0,1) circle (0.35);
    \draw[colObsLines,fill=white] (0,-1) circle (0.35);
    \node at (0,1) {$-$};
    \tenM{0}{0}{colM}
  \end{tikzpicture},\quad
  \begin{tikzpicture}[baseline={([yshift=-0.6ex]current bounding box.center)},scale=0.5]
    \vLine{0}{0.75}{0}{-0.75}
    \draw[colObsLines,fill=white] (0,0) circle (0.35);
  \end{tikzpicture}=P,\quad
  \begin{tikzpicture}[baseline={([yshift=-0.6ex]current bounding box.center)},scale=0.5]
    \vLine{0}{0.75}{0}{-0.75}
    \draw[colObsLines,fill=white] (0,0) circle (0.35);
    \node at (0,0) {$-$};
  \end{tikzpicture}=P^{-1},
\end{eqnarray} 
with the local matrix $P$ given as,
\begin{eqnarray}
  P=
  \left[\matrix{
    1 & 0 & 0 & 0 & 0 & 0 & 0 & 0 & 0\cr
    0 & 1 & 1 & 0 & 0 & 0 & 0 & 0 & 0\cr
    0 & 1 & \frac{-\vartheta}{1-\vartheta} & 0 & 0 & 0 & 0 & 0 & 0\cr
    0 & 0 & 0 & 1 & 0 & 0 & 1 & 0 & 0\cr
    0 & 0 & 0 & 0 & 1 & 1 & 0 & 1 & 1\cr
    0 & 0 & 0 & 0 & 1 & \frac{-\vartheta}{1-\vartheta} & 0 &
    1 & \frac{-\vartheta}{1-\vartheta}\cr
    0 & 0 & 0 & 1 & 0 & 0 & \frac{-\vartheta}{1-\vartheta} & 0 & 0\cr
    0 & 0 & 0 & 0 & \frac{-\vartheta}{1-\vartheta} & \frac{-\vartheta}{1-\vartheta} & 0 & 1 & 1 \cr
    0&0&0&0&\frac{-\vartheta}{1-\vartheta}&\frac{\vartheta^2}{(1-\vartheta)^2}&0&
    1&\frac{-\vartheta}{1-\vartheta}
  }\right].
\end{eqnarray}
This gives us a transformed transfer matrix
$\tilde{\mathcal{T}}_{\alpha,\vec{\gamma}}$ that has the same spectrum as
$\mathcal{T}_{\alpha,\vec{\gamma}}$,
\begin{eqnarray}
  \tilde{\mathcal{T}}_{\alpha,\vec{\gamma}}=
  \begin{tikzpicture}[baseline={([yshift=-0.6ex]current bounding box.center)},scale=0.5]
    \vLine{-0.5}{-0.5}{2.5}{-0.5}
    \vLine{-0.5}{0}{2.5}{0}
    \vleftHook{-0.5}{0}
    \vrightHook{2.5}{0}
    \foreach \x in {0,...,2}
    {
      \fill[white] (\x-0.2,-1) rectangle (\x+0.2,1);
      \vLine{\x}{-1}{\x}{1}
      \tenM{\x}{0}{colMt}
    }
  \end{tikzpicture}.
\end{eqnarray}
The transformed tensors~\eref{eq:transformedTensorsENT} satisfy Lemma 1 from
Ref.~\cite{klobas2021entanglement}. Namely, for any integer power $m\ge 1$
the following holds,
\begin{eqnarray} \label{eq:Lemma1}
  \begin{tikzpicture}[baseline={([yshift=-0.6ex]current bounding box.center)},scale=0.5]
    \foreach \y in {0,2,6,8,10}{
      \vLine{0.25}{\y}{2.75}{\y}
      \vLine{1}{\y+0.75}{1}{\y-0.75}
      \vLine{2}{\y+0.75}{2}{\y-0.75}
      \tenM{1}{\y}{colMt}
      \tenM{2}{\y}{colMt}
    }
    \node at (1,11) {\scalebox{0.8}{$v_1$}};
    \node at (2,11) {\scalebox{0.8}{$w_1$}};
    \node at (1,9) {\scalebox{0.8}{$v_2$}};
    \node at (2,9) {\scalebox{0.8}{$w_2$}};
    \node at (1,7) {\scalebox{0.8}{$v_3$}};
    \node at (2,7) {\scalebox{0.8}{$w_3$}};
    \node at (1,5) {\scalebox{0.8}{$v_4$}};
    \node at (2,5) {\scalebox{0.8}{$w_4$}};
    \node at (1.5,4.25) {$\vdots$};
    \node at (0.875,3) {\scalebox{0.78}{$v_{m\!-\!1}$}};
    \node at (2.125,3) {\scalebox{0.78}{$w_{m\!-\!1}$}};
    \node at (1,1) {\scalebox{0.8}{$v_m$}};
    \node at (2,1) {\scalebox{0.8}{$w_m$}};
    \node at (1,-1) {\scalebox{0.8}{$v_1$}};
    \node at (2,-1) {\scalebox{0.8}{$w_1$}};
  \end{tikzpicture}
  =\prod_{j=1}^{m}\delta_{w_j,v_j}
  \begin{tikzpicture}[baseline={([yshift=-0.6ex]current bounding box.center)},scale=0.5]
    \foreach \y in {0,2,6,8,10}{
      \vLine{0.25}{\y}{2.75}{\y}
      \vLine{1}{\y+0.75}{1}{\y-0.75}
      \vLine{2}{\y+0.75}{2}{\y-0.75}
      \tenM{1}{\y}{colMt}
      \tenM{2}{\y}{colMt}
    }
    \node at (1,11) {\scalebox{0.8}{$v_1$}};
    \node at (2,11) {\scalebox{0.8}{$v_1$}};
    \node at (1,9) {\scalebox{0.8}{$v_2$}};
    \node at (2,9) {\scalebox{0.8}{$v_2$}};
    \node at (1,7) {\scalebox{0.8}{$v_3$}};
    \node at (2,7) {\scalebox{0.8}{$v_3$}};
    \node at (1,5) {\scalebox{0.8}{$v_4$}};
    \node at (2,5) {\scalebox{0.8}{$v_4$}};
    \node at (1.5,4.25) {$\vdots$};
    \node at (0.875,3) {\scalebox{0.78}{$v_{m\!-\!1}$}};
    \node at (2.125,3) {\scalebox{0.78}{$v_{m\!-\!1}$}};
    \node at (1,1) {\scalebox{0.8}{$v_m$}};
    \node at (2,1) {\scalebox{0.8}{$v_m$}};
    \node at (1,-1) {\scalebox{0.8}{$v_1$}};
    \node at (2,-1) {\scalebox{0.8}{$v_1$}};
  \end{tikzpicture}.
\end{eqnarray}
The proof follows from a straightforward modification of the proof given in
Ref.~\cite{klobas2021entanglement}, and will be not repeated here.

The property~\eref{eq:Lemma1} directly implies that traces of powers of
$\tilde{\mathcal{T}}_{\alpha,\vec{\gamma}}$ (and therefore also
$\mathcal{T}_{\alpha,\vec{\gamma}}$) are given as 
\begin{eqnarray}
  \tr[\mathcal{T}_{\alpha,\vec{\gamma}}^m]=
  \tr[\tilde{\mathcal{T}}_{\alpha,\vec{\gamma}}^m]=
  \tr[\tilde{\tau}_{\alpha,\vec{\gamma}}^m],
\end{eqnarray}
where  $\tilde{\tau}_{\alpha,\vec{\gamma}}$ is a $9\times 9$ matrix given by following
matrix elements,
\begin{eqnarray}
  [\tilde{\tau}_{\alpha,\vec{\gamma}}]_{w,v}=
  \begin{tikzpicture}[baseline={([yshift=-0.6ex]current bounding box.center)},scale=0.5]
    \foreach \y in {0}{
      \vLine{0.5}{\y}{4.75}{\y}
      \vLine{0.5}{\y-0.5}{4.75}{\y-0.5}
      \vLine{6.25}{\y}{7.5}{\y}
      \vLine{6.25}{\y-0.5}{7.5}{\y-0.5}
      \vleftHook{0.5}{\y}
      \vrightHook{7.5}{\y}
      \foreach \x in {1,2,3,4,7}{
        \fill[white] (\x+0.2,\y+0.75) rectangle (\x-0.2,\y-0.75);
        \vLine{\x}{\y+0.75}{\x}{\y-0.75}
        \tenM{\x}{\y}{colMt}
      }
      \node at (5.5,\y-0.25) {$\cdots$};
    }
    \foreach \x in {1,2,3,4,7}{
      \node at (\x,1) {\scalebox{0.8}{$w$}};
      \node at (\x,-1) {\scalebox{0.8}{$v$}};
    }
  \end{tikzpicture}.
\end{eqnarray}
Explicitly, the matrix $\tilde{\tau}_{\alpha,\vec{\gamma}}$ reads as
\begin{eqnarray}\fl
  \tilde{\tau}_{\alpha,\vec{\gamma}}=
  \left[\matrix{
    (1-\vartheta)^{2\alpha}&0&0&0&0&\Gamma&(1-\vartheta)^{\alpha}&0&\Gamma\cr
    \vartheta^{\alpha}(1-\vartheta)^{\alpha}&0&0&0&0&0&\vartheta^{\alpha}&0&0\cr
    0&0&0&0&\Gamma&0&0&\Gamma&0\cr
    0&\vartheta^\alpha \Gamma&\vartheta^\alpha\Gamma&0&0&0&0&0&0\cr
    0&0&0&\vartheta^{\alpha}&0&0&0&0&0\cr
    0&0&0&(1-\vartheta)^{\alpha}&0&0&0&0&0\cr
    0&\mkern-32mu(1-\vartheta)^{\alpha}\Gamma&(1-\vartheta)^\alpha\Gamma&0&0&0&0&0&0\cr
    \vartheta^{2\alpha}&0&0&0&0&0&0&0&0\cr
    \vartheta^{\alpha}(1-\vartheta)^{\alpha}&0&0&0&0&0&0&0&0
  }\right]\mkern-4mu,
\end{eqnarray}
where $\Gamma$ is a shorthand notation for the product of $\gamma_j$,
\begin{eqnarray}
  \Gamma=\prod_{j=1}^{\alpha}\gamma_j.
\end{eqnarray}

This matrix has three non-zero eigenvalues with algebraic and
geometric multiplicities equal to $1$,
\begin{eqnarray}
  \mathrm{Spect}(\tilde{\tau}_{\alpha,\vec{\gamma}})=
  \{\lambda_{\alpha,\vec{\gamma}},\lambda^{(1)}_{\alpha,\vec{\gamma}},\lambda^{(2)}_{\alpha,\vec{\gamma}},0\},
\end{eqnarray}
which are obtained as solutions to the following cubic equation,
\begin{eqnarray}
  \label{eq:r54EquationLambda}
  x^3=\left(x(1-\vartheta)^{\alpha}+\vartheta^{\alpha} \Gamma\right)^2.
\end{eqnarray}
Here we use $\lambda_{\alpha,\vec{\gamma}}$ to denote the leading eigenvalue,
and we assume it is \emph{isolated}
\begin{eqnarray}
  \abs{\lambda_{\alpha,\vec{\gamma}}}>
  \abs{\lambda^{(1)}_{\alpha,\vec{\gamma}}},
  \abs{\lambda^{(2)}_{\alpha,\vec{\gamma}}},
\end{eqnarray}
which is not always the case. However, as long as $\Gamma$ does not take a negative real value, there is a gap between the leading and the largest subleading eigenvalue for any $\vartheta\neq 0,1$.

\section{Conclusions}\label{sec:conclusions}
In this paper we extended the approach of
Refs.~\cite{klobas2021exactrelaxation,klobas2021entanglement} to study
time-evolution of charged moments in Rule 54. We formulated the charged moments
of the two simplest charges (the total particle number, and the imbalance
between right and left movers) in terms of a space transfer matrix, which can
be understood as a deformation of the transfer-matrix capturing the dynamics of
local observables. We then found the fixed points of this deformed transfer
matrix, which are made of tensors that fulfil a set of local algebraic
relations analogous to the non-deformed case. This enabled us to fully characterize
the charged moments in the early-time regime. We showed that they decay exponentially
with a rate given in terms of a solution of a cubic equation. As was argued in
Refs.~\cite{bertini2023nonequilibrium,bertini2024dynamics}, this is a special
case of a more general non-linear equation applying for interacting integrable
systems, which can be obtained from an equation for equilibrium charged moments
upon flipping the roles of space and time.

This work leaves several open questions. Perhaps the most urgent one is to understand what are the limits of the algebraic approach to characterize space evolution. As is showed here, the fixed points can be explicitly constructed also when the space transfer matrix is deformed by applying $e^{i\beta Q^{(\pm)}_{A}}$ at time $t$, where $Q^{(\pm)}$ are the two simplest charges. The natural next step is to understand for which $Q$ is the deformed transfer matrix of Rule 54 still solvable. Does $Q$ need to be conserved, or can it be a more general operator? More generally, a key question is to find a broader class of systems that admit analogous solutions.

\ack
I thank Bruno Bertini, Pasquale Calabrese, Mario Collura, and Colin Rylands for collaborations on related topics, and Bruno Bertini for the careful reading of the manuscript. This work was supported by the Leverhulme Trust through the Early Career Fellowship No.\ ECF-2022-324. I warmly acknowledge the hospitality of the University of Ljubljana where this manuscript was finished.\\

\bibliographystyle{iopart-num-mod}
\bibliography{bibliography}

\providecommand{\newblock}{}
\begin{thebibliography}{10}
\expandafter\ifx\csname url\endcsname\relax
  \def\url#1{{\tt #1}}\fi
\expandafter\ifx\csname urlprefix\endcsname\relax\def\urlprefix{URL }\fi
\providecommand{\eprint}[2][]{\url{#2}}

\bibitem{calabrese2006time}
Calabrese P and Cardy J 2006 Time dependence of correlation functions following
  a quantum quench {\em Phys. Rev. Lett.\/}
  \href{https://doi.org/10.1103/PhysRevLett.96.136801}{{\bf 96}(13) 136801}

\bibitem{calabrese2005evolution}
Calabrese P and Cardy J 2005 Evolution of entanglement entropy in
  one-dimensional systems {\em J. Stat. Mech.: Theory Exp.\/}
  \href{https://doi.org/10.1088/1742-5468/2005/04/p04010}{{\bf 2005} P04010}

\bibitem{fagotti2008evolution}
Fagotti M and Calabrese P 2008 Evolution of entanglement entropy following a
  quantum quench: Analytic results for the {XY} chain in a transverse magnetic
  field {\em Phys. Rev. A\/}
  \href{https://doi.org/10.1103/PhysRevA.78.010306}{{\bf 78}(1) 010306}

\bibitem{alba2017entanglement}
Alba V and Calabrese P 2017 Entanglement and thermodynamics after a quantum
  quench in integrable systems {\em Proc. Natl. Acad. Sci. U.S.A.\/}
  \href{https://doi.org/10.1073/pnas.1703516114}{{\bf 114} 7947--7951}

\bibitem{alba2018entanglement}
Alba V and Calabrese P 2018 Entanglement dynamics after quantum quenches in
  generic integrable systems {\em SciPost Phys.\/}
  \href{https://doi.org/10.21468/SciPostPhys.4.3.017}{{\bf 4}(3) 17}

\bibitem{alba2019entanglement}
Alba V, Bertini B and Fagotti M 2019 Entanglement evolution and generalised
  hydrodynamics: {I}nteracting integrable systems {\em SciPost Phys.\/}
  \href{https://doi.org/10.21468/SciPostPhys.7.1.005}{{\bf 7}(1) 5}

\bibitem{lagnese2022entanglement}
Lagnese G, Calabrese P and Piroli L 2022 Entanglement dynamics of thermofield
  double states in integrable models {\em J. Phys. A: Math. Theor.\/}
  \href{https://doi.org/10.1088/1751-8121/ac646b}{{\bf 55} 214003}

\bibitem{laeuchli2008spreading}
L\"auchli A~M and Kollath C 2008 Spreading of correlations and entanglement
  after a quench in the one-dimensional {B}ose--{H}ubbard model {\em J. Stat.
  Mech.: Theory Exp.\/}
  \href{https://doi.org/10.1088/1742-5468/2008/05/p05018}{{\bf 2008} P05018}

\bibitem{kim2013ballistic}
Kim H and Huse D~A 2013 Ballistic spreading of entanglement in a diffusive
  nonintegrable system {\em Phys. Rev. Lett.\/}
  \href{https://doi.org/10.1103/PhysRevLett.111.127205}{{\bf 111}(12) 127205}

\bibitem{liu2014entanglement}
Liu H and Suh S~J 2014 Entanglement tsunami: Universal scaling in holographic
  thermalization {\em Phys. Rev. Lett.\/}
  \href{https://doi.org/10.1103/PhysRevLett.112.011601}{{\bf 112}(1) 011601}

\bibitem{asplund2015entanglement}
Asplund C~T, Bernamonti A, Galli F and Hartman T 2015 Entanglement scrambling
  in 2d conformal field theory {\em J. High Energy Phys.\/}
  \href{https://doi.org/10.1007/JHEP09(2015)110}{{\bf 2015} 110}

\bibitem{nahum2017quantum}
Nahum A, Ruhman J, Vijay S and Haah J 2017 Quantum entanglement growth under
  random unitary dynamics {\em Phys. Rev. X\/}
  \href{https://doi.org/10.1103/PhysRevX.7.031016}{{\bf 7}(3) 031016}

\bibitem{pal2018entangling}
Pal R and Lakshminarayan A 2018 Entangling power of time-evolution operators in
  integrable and nonintegrable many-body systems {\em Phys. Rev. B\/}
  \href{https://doi.org/10.1103/PhysRevB.98.174304}{{\bf 98}(17) 174304}

\bibitem{bertini2019entanglement}
Bertini B, Kos P and Prosen T 2019 Entanglement spreading in a minimal model of
  maximal many-body quantum chaos {\em Phys. Rev. X\/}
  \href{https://doi.org/10.1103/PhysRevX.9.021033}{{\bf 9} 021033}

\bibitem{gopalakrishnan2019unitary}
Gopalakrishnan S and Lamacraft A 2019 Unitary circuits of finite depth and
  infinite width from quantum channels {\em Phys. Rev. B\/}
  \href{https://doi.org/10.1103/PhysRevB.100.064309}{{\bf 100}(6) 064309}

\bibitem{piroli2020exact}
Piroli L, Bertini B, Cirac J~I and Prosen T 2020 Exact dynamics in dual-unitary
  quantum circuits {\em Phys. Rev. B\/}
  \href{https://doi.org/10.1103/PhysRevB.101.094304}{{\bf 101}(9) 094304}

\bibitem{zhou2020entanglement}
Zhou T and Nahum A 2020 Entanglement membrane in chaotic many-body systems {\em
  Phys. Rev. X\/} \href{https://doi.org/10.1103/PhysRevX.10.031066}{{\bf 10}(3)
  031066}

\bibitem{amico2008entanglement}
Amico L, Fazio R, Osterloh A and Vedral V 2008 Entanglement in many-body
  systems {\em Rev. Mod. Phys.\/}
  \href{https://doi.org/10.1103/RevModPhys.80.517}{{\bf 80}(2) 517--576}

\bibitem{horodecki2009quantum}
Horodecki R, Horodecki P, Horodecki M and Horodecki K 2009 Quantum entanglement
  {\em Rev. Mod. Phys.\/} \href{https://doi.org/10.1103/RevModPhys.81.865}{{\bf
  81}(2) 865--942}

\bibitem{laflorencie2014spin}
Laflorencie N and Rachel S 2014 Spin-resolved entanglement spectroscopy of
  critical spin chains and {L}uttinger liquids {\em J. Stat. Mech.: Theory
  Exp.\/} \href{https://doi.org/10.1088/1742-5468/2014/11/P11013}{{\bf 2014}
  P11013}

\bibitem{goldstein2018symmetry}
Goldstein M and Sela E 2018 Symmetry-resolved entanglement in many-body systems
  {\em Phys. Rev. Lett.\/}
  \href{https://doi.org/10.1103/PhysRevLett.120.200602}{{\bf 120}(20) 200602}

\bibitem{xavier2018equipartition}
Xavier J~C, Alcaraz F~C and Sierra G 2018 Equipartition of the entanglement
  entropy {\em Phys. Rev. B\/}
  \href{https://doi.org/10.1103/PhysRevB.98.041106}{{\bf 98}(4) 041106}

\bibitem{bonsignori2019symmetry}
Bonsignori R, Ruggiero P and Calabrese P 2019 Symmetry resolved entanglement in
  free fermionic systems {\em J. Phys. A: Math. Theor.\/}
  \href{https://doi.org/10.1088/1751-8121/ab4b77}{{\bf 52} 475302}

\bibitem{murciano2020entanglement}
Murciano S, Di~Giulio G and Calabrese P 2020 Entanglement and symmetry
  resolution in two dimensional free quantum field theories {\em J. High Energy
  Phys.\/} \href{https://doi.org/10.1007/JHEP08%282020%29073}{{\bf 2020} 1--42}

\bibitem{ares2023entanglement}
Ares F, Murciano S and Calabrese P 2023 Entanglement asymmetry as a probe of
  symmetry breaking {\em Nat. Commun.\/}
  \href{https://doi.org/10.1038/s41467-023-37747-8}{{\bf 14} 2036}

\bibitem{ares2023lack}
Ares F, Murciano S, Vernier E and Calabrese P 2023 Lack of symmetry restoration
  after a quantum quench: An entanglement asymmetry study {\em SciPost Phys.\/}
  \href{https://doi.org/10.21468/SciPostPhys.15.3.089}{{\bf 15} 089}

\bibitem{capizzi2023entanglement}
Capizzi L and Mazzoni M 2023 Entanglement asymmetry in the ordered phase of
  many-body systems: The {I}sing field theory {\em J. High Energy Phys.\/}
  \href{https://doi.org/10.1007/JHEP12(2023)144}{{\bf 2023} 1--36}

\bibitem{capizzi2023universal}
Capizzi L and Vitale V 2024 A universal formula for the entanglement asymmetry
  of matrix product states \textit{Preprint}
  \href{http://arxiv.org/abs/2310.01962}{{ arXiv:2310.01962}}

\bibitem{khor2023confinement}
Khor B~J~J, Kürkçüoglu D~M, Hobbs T~J, Perdue G~N and Klich I 2023
  Confinement and kink entanglement asymmetry on a quantum {I}sing chain
  \textit{Preprint} \href{http://arxiv.org/abs/2312.08601}{{ arXiv:2312.08601}}

\bibitem{bertini2023nonequilibrium}
Bertini B, Calabrese P, Collura M, Klobas K and Rylands C 2023 Nonequilibrium
  full counting statistics and symmetry-resolved entanglement from space-time
  duality {\em Phys. Rev. Lett.\/}
  \href{https://doi.org/10.1103/PhysRevLett.131.140401}{{\bf 131}(14) 140401}

\bibitem{bertini2024dynamics}
Bertini B, Klobas K, Collura M, Calabrese P and Rylands C 2024 Dynamics of
  charge fluctuations from asymmetric initial states {\em Phys. Rev. B\/}
  \href{https://doi.org/10.1103/PhysRevB.109.184312}{{\bf 109} 184312}

\bibitem{murciano2024entanglement}
Murciano S, Ares F, Klich I and Calabrese P 2024 Entanglement asymmetry and
  quantum {M}pemba effect in the {XY} spin chain {\em J. Stat. Mech.: Theory
  Exp.\/} \href{https://doi.org/10.1088/1742-5468/ad17b4}{{\bf 2024} 013103}

\bibitem{caceffo2024entangled}
Caceffo F, Murciano S and Alba V 2024 Entangled multiplets, asymmetry, and
  quantum {M}pemba effect in dissipative systems {\em J. Stat. Mech.: Theory
  Exp.\/} \href{https://doi.org/10.1088/1742-5468/ad4537}{{\bf 2024} 063103}

\bibitem{ferro2024nonequilibrium}
Ferro F, Ares F and Calabrese P 2024 Non-equilibrium entanglement asymmetry for
  discrete groups: The example of the {XY} spin chain {\em J. Stat. Mech.:
  Theory Exp.\/} \href{https://doi.org/10.1088/1742-5468/ad138f}{{\bf 2024}
  023101}

\bibitem{rylands2024microscopic}
Rylands C, Klobas K, Ares F, Calabrese P, Murciano S and Bertini B 2024
  Microscopic origin of the quantum {M}pemba effect in integrable systems {\em
  Phys. Rev. Lett.\/}
  \href{https://doi.org/10.1103/PhysRevLett.133.010401}{{\bf 133}(1) 010401}

\bibitem{chen2024renyi}
Chen M and Chen H~H 2024 R\'enyi entanglement asymmetry in ($1+1$)-dimensional
  conformal field theories {\em Phys. Rev. D\/}
  \href{https://doi.org/10.1103/PhysRevD.109.065009}{{\bf 109}(6) 065009}

\bibitem{banerjee2024symmetry}
Banerjee A, Basu R, Bhattacharyya A and Chakrabarti N 2024 Symmetry resolution
  in non-{L}orentzian field theories {\em J. High Energy Phys.\/}
  \href{https://doi.org/10.1007/JHEP06(2024)121}{{\bf 2024} 1--36}

\bibitem{lukin2019probing}
Lukin A, Rispoli M, Schittko R, Tai M~E, Kaufman A~M, Choi S, Khemani V,
  Léonard J and Greiner M 2019 Probing entanglement in a many-body localized
  system {\em Science\/} \href{https://doi.org/10.1126/science.aau0818}{{\bf
  364} 256--260}

\bibitem{azses2020identification}
Azses D, Haenel R, Naveh Y, Raussendorf R, Sela E and Dalla~Torre E~G 2020
  Identification of symmetry-protected topological states on noisy quantum
  computers {\em Phys. Rev. Lett.\/}
  \href{https://doi.org/10.1103/PhysRevLett.125.120502}{{\bf 125}(12) 120502}

\bibitem{neven2021symmetry}
Neven A, Carrasco J, Vitale V, Kokail C, Elben A, Dalmonte M, Calabrese P,
  Zoller P, Vermersch B, Kueng R {\em et~al.\/} 2021 Symmetry-resolved
  entanglement detection using partial transpose moments {\em Npj Quantum
  Inf.\/} \href{https://doi.org/10.1038/s41534-021-00487-y}{{\bf 7} 1--12}

\bibitem{vitale2022symmetry}
Vitale V, Elben A, Kueng R, Neven A, Carrasco J, Kraus B, Zoller P, Calabrese
  P, Vermersch B and Dalmonte M 2022 Symmetry-resolved dynamical purification
  in synthetic quantum matter {\em SciPost Phys.\/}
  \href{https://doi.org/10.21468/SciPostPhys.12.3.106}{{\bf 12} 106}

\bibitem{rath2023entanglement}
Rath A, Vitale V, Murciano S, Votto M, Dubail J, Kueng R, Branciard C,
  Calabrese P and Vermersch B 2023 Entanglement barrier and its symmetry
  resolution: Theory and experimental observation {\em PRX Quantum\/}
  \href{https://doi.org/10.1103/PRXQuantum.4.010318}{{\bf 4}(1) 010318}

\bibitem{joshi2024observing}
Joshi L~K, Franke J, Rath A, Ares F, Murciano S, Kranzl F, Blatt R, Zoller P,
  Vermersch B, Calabrese P {\em et~al.\/} 2024 Observing the quantum {M}pemba
  effect in quantum simulations {\em Phys. Rev. Lett.\/}
  \href{https://doi.org/10.1103/PhysRevLett.133.010402}{{\bf 133} 010402}

\bibitem{bertini2022growth}
Bertini B, Klobas K, Alba V, Lagnese G and Calabrese P 2022 Growth of {R}\'enyi
  entropies in interacting integrable models and the breakdown of the
  quasiparticle picture {\em Phys. Rev. X\/}
  \href{https://doi.org/10.1103/PhysRevX.12.031016}{{\bf 12}(3) 031016}

\bibitem{caux2016quench}
Caux J~S 2016 The quench action {\em J. Stat. Mech.: Theory Exp.\/}
  \href{https://doi.org/10.1088/1742-5468/2016/06/064006}{{\bf 2016} 064006}

\bibitem{takahashi1999thermodynamics}
Takahashi M 1999 {\em Thermodynamics of One-Dimensional Solvable Models\/}
  (Cambridge University Press)

\bibitem{bobenko1993two}
Bobenko A, Bordemann M, Gunn C and Pinkall U 1993 On two integrable cellular
  automata {\em Commun. Math. Phys.\/}
  \href{https://doi.org/https://doi.org/10.1007/BF02097234}{{\bf 158} 127--134}

\bibitem{friedman2019integrable}
Friedman A~J, Gopalakrishnan S and Vasseur R 2019 Integrable many-body quantum
  {F}loquet-{T}houless pumps {\em Phys. Rev. Lett.\/}
  \href{https://doi.org/10.1103/PhysRevLett.123.170603}{{\bf 123}(17) 170603}

\bibitem{gombor2024integrable}
Gombor T and Pozsgay B 2024 Integrable deformations of superintegrable quantum
  circuits {\em SciPost Phys.\/}
  \href{https://doi.org/10.21468/SciPostPhys.16.4.114}{{\bf 16} 114}

\bibitem{prosen2016integrability}
Prosen T and Mej{\'{\i}}a-Monasterio C 2016 Integrability of a deterministic
  cellular automaton driven by stochastic boundaries {\em J. Phys. A: Math.
  Theor.\/} \href{https://doi.org/10.1088/1751-8113/49/18/185003}{{\bf 49}
  185003}

\bibitem{prosen2017exact}
Prosen T and Bu{\v{c}}a B 2017 Exact matrix product decay modes of a boundary
  driven cellular automaton {\em J. Phys. A: Math. Theor.\/}
  \href{https://doi.org/10.1088/1751-8121/aa85a3}{{\bf 50} 395002}

\bibitem{inoue2018two}
Inoue A and Takesue S 2018 Two extensions of exact nonequilibrium steady states
  of a boundary-driven cellular automaton {\em J. Phys. A: Math. Theor.\/}
  \href{https://doi.org/10.1088/1751-8121/aadc29}{{\bf 51} 425001}

\bibitem{klobas2019time}
Klobas K, Medenjak M, Prosen T and Vanicat M 2019 Time-dependent matrix product
  ansatz for interacting reversible dynamics {\em Commun. Math. Phys.\/}
  \href{https://doi.org/10.1007/s00220-019-03494-5}{{\bf 371} 651--688}

\bibitem{buca2019exact}
Bu{\v c}a B, Garrahan J~P, Prosen T and Vanicat M 2019 Exact large deviation
  statistics and trajectory phase transition of a deterministic boundary driven
  cellular automaton {\em Phys. Rev. E\/}
  \href{https://doi.org/10.1103/PhysRevE.100.020103}{{\bf 100}(2) 020103(R)}

\bibitem{klobas2020matrix}
Klobas K, Vanicat M, Garrahan J~P and Prosen T 2020 Matrix product state of
  multi-time correlations {\em J. Phys. A: Math. Theor.\/}
  \href{https://doi.org/10.1088/1751-8121/ab8c62}{{\bf 53} 335001}

\bibitem{klobas2020space}
Klobas K and Prosen T 2020 Space-like dynamics in a reversible cellular
  automaton {\em SciPost Phys. Core\/}
  \href{https://doi.org/10.21468/SciPostPhysCore.2.2.010}{{\bf 2}(2) 10}

\bibitem{gopalakrishnan2018operator}
Gopalakrishnan S 2018 Operator growth and eigenstate entanglement in an
  interacting integrable {F}loquet system {\em Phys. Rev. B\/}
  \href{https://doi.org/10.1103/PhysRevB.98.060302}{{\bf 98}(6) 060302(R)}

\bibitem{gopalakrishnan2018hydrodynamics}
Gopalakrishnan S, Huse D~A, Khemani V and Vasseur R 2018 Hydrodynamics of
  operator spreading and quasiparticle diffusion in interacting integrable
  systems {\em Phys. Rev. B\/}
  \href{https://doi.org/10.1103/PhysRevB.98.220303}{{\bf 98}(22) 220303(R)}

\bibitem{alba2019operator}
Alba V, Dubail J and Medenjak M 2019 Operator entanglement in interacting
  integrable quantum systems: The case of the {R}ule 54 chain {\em Phys. Rev.
  Lett.\/} \href{https://doi.org/10.1103/PhysRevLett.122.250603}{{\bf 122}(25)
  250603}

\bibitem{buca2021rule}
Bu\v{c}a B, Klobas K and Prosen T 2021 Rule 54: Exactly solvable model of
  nonequilibrium statistical mechanics {\em J. Stat. Mech.: Theory Exp.\/}
  \href{https://doi.org/10.1088/1742-5468/ac096b}{{\bf 2021} 074001}

\bibitem{klobas2021exact}
Klobas K, Bertini B and Piroli L 2021 Exact thermalization dynamics in the
  ``{R}ule 54'' quantum cellular automaton {\em Phys. Rev. Lett.\/}
  \href{https://doi.org/10.1103/PhysRevLett.126.160602}{{\bf 126}(16) 160602}

\bibitem{klobas2021exactrelaxation}
Klobas K and Bertini B 2021 Exact relaxation to {Gibbs} and non-equilibrium
  steady states in the quantum cellular automaton {Rule} 54 {\em SciPost
  Phys.\/} \href{https://doi.org/10.21468/SciPostPhys.11.6.106}{{\bf 11}(6)
  106}

\bibitem{klobas2021entanglement}
Klobas K and Bertini B 2021 Entanglement dynamics in {R}ule 54: Exact results
  and quasiparticle picture {\em SciPost Phys.\/}
  \href{https://doi.org/10.21468/SciPostPhys.11.6.107}{{\bf 11}(6) 107}

\bibitem{banuls2009matrix}
Ba\~nuls M~C, Hastings M~B, Verstraete F and Cirac J~I 2009 Matrix product
  states for dynamical simulation of infinite chains {\em Phys. Rev. Lett.\/}
  \href{https://doi.org/10.1103/PhysRevLett.102.240603}{{\bf 102}(24) 240603}

\bibitem{muller2012tensor}
M{\"u}ller-Hermes A, Cirac J~I and Ba{\~n}uls M~C 2012 Tensor network
  techniques for the computation of dynamical observables in one-dimensional
  quantum spin systems {\em New J. Phys.\/}
  \href{https://doi.org/10.1088/1367-2630/14/7/075003}{{\bf 14} 075003}

\bibitem{hastings2015connecting}
Hastings M~B and Mahajan R 2015 Connecting entanglement in time and space:
  Improving the folding algorithm {\em Phys. Rev. A\/}
  \href{https://doi.org/10.1103/PhysRevA.91.032306}{{\bf 91}(3) 032306}

\bibitem{bertini2018exact}
Bertini B, Kos P and Prosen T 2018 Exact spectral form factor in a minimal
  model of many-body quantum chaos {\em Phys. Rev. Lett.\/}
  \href{https://doi.org/10.1103/PhysRevLett.121.264101}{{\bf 121}(26) 264101}

\bibitem{bertini2019exact}
Bertini B, Kos P and Prosen T 2019 Exact correlation functions for dual-unitary
  lattice models in $1+1$ dimensions {\em Phys. Rev. Lett.\/}
  \href{https://doi.org/10.1103/PhysRevLett.123.210601}{{\bf 123}(21) 210601}

\bibitem{lerose2021influence}
Lerose A, Sonner M and Abanin D~A 2021 Influence matrix approach to many-body
  {F}loquet dynamics {\em Phys. Rev. X\/}
  \href{https://doi.org/10.1103/PhysRevX.11.021040}{{\bf 11} 021040}

\bibitem{lerose2021scaling}
Lerose A, Sonner M and Abanin D~A 2021 Scaling of temporal entanglement in
  proximity to integrability {\em Phys. Rev. B\/}
  \href{https://doi.org/10.1103/PhysRevB.104.035137}{{\bf 104} 035137}

\bibitem{sonner2021influence}
Sonner M, Lerose A and Abanin D~A 2021 Influence functional of many-body
  systems: Temporal entanglement and matrix-product state representation {\em
  Ann. Physics\/} \href{https://doi.org/10.1016/j.aop.2021.168677}{{\bf 435}
  168677}

\bibitem{ippoliti2021postselectionfree}
Ippoliti M and Khemani V 2021 Postselection-free entanglement dynamics via
  spacetime duality {\em Phys. Rev. Lett.\/}
  \href{https://doi.org/10.1103/PhysRevLett.126.060501}{{\bf 126}(6) 060501}

\bibitem{bertini2022entanglement}
Bertini B, Klobas K and Lu T~C 2022 Entanglement negativity and mutual
  information after a quantum quench: Exact link from space-time duality {\em
  Phys. Rev. Lett.\/}
  \href{https://doi.org/10.1103/PhysRevLett.129.140503}{{\bf 129}(14) 140503}

\bibitem{haegeman2017diagonalizing}
Haegeman J and Verstraete F 2017 Diagonalizing transfer matrices and matrix
  product operators: A medley of exact and computational methods {\em Annu.
  Rev. Condens. Matter Phys.\/}
  \href{https://doi.org/10.1146/annurev-conmatphys-031016-025507}{{\bf 8}
  355--406}

\end{thebibliography}
\appendix

\section{Spectral properties of space transfer matrices}\label{sec:uniquenessFPs}
To find the spectrum of the space-transfer matrices we first note that by
definition we have for any $L\ge 1$
\begin{eqnarray}
  \tr[{\mathbb{W}_{\beta,t}^{(\pm)\, L}}] = 
  \sbra{\Psi_t}e^{i \beta Q^{(\pm)}}\sket{\Psi_t} =
  \sbra{\Psi_0}e^{i \beta Q^{(\pm)}}\sket{\Psi_0} =
  \tr[{\mathbb{W}_{\beta,0}^{(\pm)\, L}}],
\end{eqnarray}
where the r.h.s.\ corresponds to the system defined on $2L$ sites with periodic
boundary conditions, in the second equality we used that $Q^{(\pm)}$ is
conserved under time-evolution, and the last equality again follows from the definition
of the transfer matrix. Transfer matrices $\mathbb{W}_{\beta,0}^{(\pm)}$ act
only on the auxiliary space of the $e^{i\beta Q}$ MPO, and take the following form,
\begin{eqnarray}\fl
  \mathbb{W}_{\beta,0}^{(+)}=
  \begin{tikzpicture}[baseline={([yshift=-0.6ex]current bounding box.center)},scale=0.55]
    \def\Y{0}

    \gridLine{1}{0}{1}{2*\Y+2}
    \gridLine{2}{0}{2}{2*\Y+2}

    \vmpsV{1}{0}{colIst}{colLines}
    \vmpsW{2}{0}{colIst}{colLines}
    \vmpsV{1}{2*\Y+2}{colIstC}{colLines}
    \vmpsW{2}{2*\Y+2}{colIstC}{colLines}

    \vLine{0.25}{\Y+1}{2.75}{\Y+1}
    \vmpsNS{1}{\Y+1}{colObs}{colObsLines}
    \vmpsNS{2}{\Y+1}{colObs}{colObsLines}
  \end{tikzpicture}=
  \left[\matrix{
    1-\vartheta & \vartheta & 0 \cr
    e^{2i\beta}(1-\vartheta) & e^{2i\beta}\vartheta & 0 \cr
    (1-\vartheta) & \vartheta & 0 
  }\right],\quad
  \mathbb{W}_{\beta,0}^{(-)}=
  \begin{tikzpicture}[baseline={([yshift=-0.6ex]current bounding box.center)},scale=0.55]
    \def\Y{0}

    \gridLine{1}{0}{1}{2*\Y+2}
    \gridLine{2}{0}{2}{2*\Y+2}

    \vmpsV{1}{0}{colIst}{colLines}
    \vmpsW{2}{0}{colIst}{colLines}
    \vmpsV{1}{2*\Y+2}{colIstC}{colLines}
    \vmpsW{2}{2*\Y+2}{colIstC}{colLines}

    \vLine{0.25}{\Y+1}{2.75}{\Y+1}
    \vmpsV{1}{\Y+1}{colObs}{colObsLines}
    \vmpsW{2}{\Y+1}{colObs}{colObsLines}
  \end{tikzpicture}=
  \left[\matrix{
    1-\vartheta & \vartheta \cr
    1-\vartheta & \vartheta
  }\right].
\end{eqnarray}
Both these matrices have exactly one non-zero eigenvalue, which we denote by 
$\Lambda_{\beta}^{(\pm)}$, and are
\begin{eqnarray}
  \Lambda_{\beta}^{(+)}=1-\vartheta+e^{2i\beta}\vartheta,\qquad
  \Lambda_{\beta}^{(-)}=1.
\end{eqnarray}

Note that the matrices $W_{\beta,0}^{(\pm)}$ have trivial Jordan structure, and therefore
\begin{eqnarray}\label{eq:TM0eigen}
  \frac{1}{\Lambda_{\beta}^{(\pm)}}W_{\beta,0}^{(\pm)}=
  \frac{\sketbra{r_{\beta,0}^{(\pm)}}{l_{\beta,0}^{(\pm)}}}
  {\sbraket{r_{\beta,0}^{(\pm)}}{l_{\beta,0}^{(\pm)}}}.
\end{eqnarray}
This ceases to be the case for $t>0$, when the Jordan structure of the transfer
matrix becomes richer, but using the conservation of $Q^{(\pm)}$ together with 
locality of time-evolution and~\eref{eq:TM0eigen} one can show that we have
\begin{eqnarray}
  \left.\left(\frac{1}{\Lambda_{\beta}^{(\pm)}}W_{\beta,0}^{(\pm)}\right)^x
  \right|_{x > 2t+1}=
  \frac{\sketbra{r_{\beta,0}^{(\pm)}}{l_{\beta,0}^{(\pm)}}}
  {\sbraket{r_{\beta,0}^{(\pm)}}{l_{\beta,0}^{(\pm)}}}.
\end{eqnarray}

\section{Fixed-point tensors}\label{sec:tensorsOther}
\subsection{Auxiliary two-step tensors}\label{sec:tensorC}
Here we report the remaining tensor needed for
Eqs.~\eref{eq:localRelationsLeftBulkBottom},~\eref{eq:localRelationsLeftBeta},
and~\eref{eq:localRelationsRightBeta} to hold. The blue tensors take the
following form,
\begin{eqnarray}\fl
  \eqalign{
    \begin{tikzpicture}[baseline={([yshift=-0.6ex]current bounding box.center)},scale=0.5]
      \mpsWire{0}{-0.625}{0}{1.625}
      \gridLine{-0.55}{0}{0.55}{0}
      \gridLine{-0.55}{1}{0.55}{1}
      \mpsC{0}{0}{1}{colMPS}
      \node at (-0.875,1) {\scalebox{0.8}{$0$}};
      \node at (0.875,1) {\scalebox{0.8}{$0$}};
      \node at (-0.875,0) {\scalebox{0.8}{$0$}};
      \node at (0.875,0) {\scalebox{0.8}{$0$}};
    \end{tikzpicture}=
    \left[\matrix{
      (1-\vartheta)^2 & (1-\vartheta)^2 & -(1-\vartheta)^2\\ 
      (1-\vartheta)\vartheta & (1-\vartheta)\vartheta & -(1-\vartheta)\vartheta\\ 
      (1-\vartheta)\vartheta & -\vartheta^2 & \vartheta^2
    }\right],\qquad
    &\begin{tikzpicture}[baseline={([yshift=-0.6ex]current bounding box.center)},scale=0.5]
      \mpsWire{0}{-0.625}{0}{1.625}
      \gridLine{-0.55}{0}{0.55}{0}
      \gridLine{-0.55}{1}{0.55}{1}
      \mpsC{0}{0}{1}{colMPS}
      \node at (-0.875,1) {\scalebox{0.8}{$1$}};
      \node at (0.875,1) {\scalebox{0.8}{$1$}};
      \node at (-0.875,0) {\scalebox{0.8}{$0$}};
      \node at (0.875,0) {\scalebox{0.8}{$0$}};
    \end{tikzpicture}=
    \left[\matrix{
      \vartheta & 0 & 0\\ 
      0 & \vartheta & (1-\vartheta)\\
      0 & 0 & 0
    }\right],\mkern-32mu \\
    \begin{tikzpicture}[baseline={([yshift=-0.6ex]current bounding box.center)},scale=0.5]
      \mpsWire{0}{-0.625}{0}{1.625}
      \gridLine{-0.55}{0}{0.55}{0}
      \gridLine{-0.55}{1}{0.55}{1}
      \mpsC{0}{0}{1}{colMPS}
      \node at (-0.875,1) {\scalebox{0.8}{$0$}};
      \node at (0.875,1) {\scalebox{0.8}{$0$}};
      \node at (-0.875,0) {\scalebox{0.8}{$0$}};
      \node at (0.875,0) {\scalebox{0.8}{$1$}};
    \end{tikzpicture}=
    \left[\matrix{
      0 & (1-\vartheta)^2 & -(1-\vartheta)^2\\ 
      0 & (1-\vartheta)\vartheta & -(1-\vartheta)\vartheta\\ 
      0 & -\vartheta^2 & \vartheta^2
    }\right],
    &\begin{tikzpicture}[baseline={([yshift=-0.6ex]current bounding box.center)},scale=0.5]
      \mpsWire{0}{-0.625}{0}{1.625}
      \gridLine{-0.55}{0}{0.55}{0}
      \gridLine{-0.55}{1}{0.55}{1}
      \mpsC{0}{0}{1}{colMPS}
      \node at (-0.875,1) {\scalebox{0.8}{$1$}};
      \node at (0.875,1) {\scalebox{0.8}{$1$}};
      \node at (-0.875,0) {\scalebox{0.8}{$0$}};
      \node at (0.875,0) {\scalebox{0.8}{$1$}};
    \end{tikzpicture}=
    \left[\matrix{
      \vartheta & 0 & 0\\ 
      0 & 0 & 0\\
      0 & 0 & 0
    }\right],\\
    \begin{tikzpicture}[baseline={([yshift=-0.6ex]current bounding box.center)},scale=0.5]
      \mpsWire{0}{-0.625}{0}{1.625}
      \gridLine{-0.55}{0}{0.55}{0}
      \gridLine{-0.55}{1}{0.55}{1}
      \mpsC{0}{0}{1}{colMPS}
      \node at (-0.875,1) {\scalebox{0.8}{$0$}};
      \node at (0.875,1) {\scalebox{0.8}{$0$}};
      \node at (-0.875,0) {\scalebox{0.8}{$1$}};
      \node at (0.875,0) {\scalebox{0.8}{$1$}};
    \end{tikzpicture}=
    \left[\matrix{
      0 & 1-\vartheta & 0\\ 
      0 & \vartheta & 0\\ 
      0 & 0 & \vartheta
    }\right],
    &\begin{tikzpicture}[baseline={([yshift=-0.6ex]current bounding box.center)},scale=0.5]
      \mpsWire{0}{-0.625}{0}{1.625}
      \gridLine{-0.55}{0}{0.55}{0}
      \gridLine{-0.55}{1}{0.55}{1}
      \mpsC{0}{0}{1}{colMPS}
      \node at (-0.875,1) {\scalebox{0.8}{$1$}};
      \node at (0.875,1) {\scalebox{0.8}{$1$}};
      \node at (-0.875,0) {\scalebox{0.8}{$1$}};
      \node at (0.875,0) {\scalebox{0.8}{$1$}};
    \end{tikzpicture}=
    \left[\matrix{
      \vartheta & 0 & 0\\ 
      1-\vartheta & 0 & 0\\
      0 & 0 & 0
    }\right],\\
    \begin{tikzpicture}[baseline={([yshift=-0.6ex]current bounding box.center)},scale=0.5]
      \mpsWire{0}{-0.625}{0}{1.625}
      \gridLine{-0.55}{0}{0.55}{0}
      \gridLine{-0.55}{1}{0.55}{1}
      \mpsC{0}{0}{1}{colMPS}
      \node at (-0.875,1) {\scalebox{0.8}{$0$}};
      \node at (0.875,1) {\scalebox{0.8}{$1$}};
      \node at (-0.875,0) {\scalebox{0.8}{$0$}};
      \node at (0.875,0) {\scalebox{0.8}{$1$}};
    \end{tikzpicture}=
    \left[\matrix{
      0 & 1-\vartheta & 0\\ 
      0 & \vartheta & 0\\ 
      0 & 0 & \vartheta
    }\right],
    &\begin{tikzpicture}[baseline={([yshift=-0.6ex]current bounding box.center)},scale=0.5]
      \mpsWire{0}{-0.625}{0}{1.625}
      \gridLine{-0.55}{0}{0.55}{0}
      \gridLine{-0.55}{1}{0.55}{1}
      \mpsC{0}{0}{1}{colMPS}
      \node at (-0.875,1) {\scalebox{0.8}{$s$}};
      \node at (0.875,1) {\scalebox{0.8}{$\mkern18mu 1\!-\!s$}};
      \node at (-0.875,0) {\scalebox{0.8}{$b$}};
      \node at (0.875,0) {\scalebox{0.8}{$b$}};
    \end{tikzpicture}=0,\\
    \begin{tikzpicture}[baseline={([yshift=-0.6ex]current bounding box.center)},scale=0.5]
      \mpsWire{0}{-0.625}{0}{1.625}
      \gridLine{-0.55}{0}{0.55}{0}
      \gridLine{-0.55}{1}{0.55}{1}
      \mpsC{0}{0}{1}{colMPS}
      \node at (-0.875,1) {\scalebox{0.8}{$0$}};
      \node at (0.875,1) {\scalebox{0.8}{$1$}};
      \node at (-0.875,0) {\scalebox{0.8}{$1$}};
      \node at (0.875,0) {\scalebox{0.8}{$0$}};
    \end{tikzpicture}=
    \left[\matrix{
      0 & 0 & 0\\ 
      0 & \vartheta^2 & (1-\vartheta)\vartheta\\ 
      0 & \vartheta^2 & (1-\vartheta)\vartheta\\ 
    }\right],
    &\begin{tikzpicture}[baseline={([yshift=-0.6ex]current bounding box.center)},scale=0.5]
      \mpsWire{0}{-0.625}{0}{1.625}
      \gridLine{-0.55}{0}{0.55}{0}
      \gridLine{-0.55}{1}{0.55}{1}
      \mpsC{0}{0}{1}{colMPS}
      \node at (-0.875,1) {\scalebox{0.8}{$s_1$}};
      \node at (0.875,1) {\scalebox{0.8}{$b_1$}};
      \node at (-0.875,0) {\scalebox{0.8}{$s_2$}};
      \node at (0.875,0) {\scalebox{0.8}{$b_2$}};
    \end{tikzpicture}=
    \begin{tikzpicture}[baseline={([yshift=-0.6ex]current bounding box.center)},scale=0.5]
      \mpsWire{0}{-0.625}{0}{1.625}
      \gridLine{-0.55}{0}{0.55}{0}
      \gridLine{-0.55}{1}{0.55}{1}
      \mpsC{0}{0}{1}{colMPS}
      \node at (-0.875,1) {\scalebox{0.8}{$b_1$}};
      \node at (0.875,1) {\scalebox{0.8}{$s_1$}};
      \node at (-0.875,0) {\scalebox{0.8}{$b_2$}};
      \node at (0.875,0) {\scalebox{0.8}{$s_2$}};
    \end{tikzpicture}.}
  \end{eqnarray}
  The orange tensors are very similar, with the majority of the tensors being the same as blue,
  \begin{eqnarray}
    \left.\begin{tikzpicture}[baseline={([yshift=-0.6ex]current bounding box.center)},scale=0.5]
      \mpsWire{0}{-0.625}{0}{1.625}
      \gridLine{-0.55}{0}{0.55}{0}
      \gridLine{-0.55}{1}{0.55}{1}
      \mpsC{0}{0}{1}{colMPSBeta}
      \node at (-0.875,1) {\scalebox{0.8}{$s_1$}};
      \node at (0.875,1) {\scalebox{0.8}{$b_1$}};
      \node at (-0.875,0) {\scalebox{0.8}{$s_2$}};
      \node at (0.875,0) {\scalebox{0.8}{$b_2$}};
    \end{tikzpicture}\right|_{s_1+s_2<2}=
    \begin{tikzpicture}[baseline={([yshift=-0.6ex]current bounding box.center)},scale=0.5]
      \mpsWire{0}{-0.625}{0}{1.625}
      \gridLine{-0.55}{0}{0.55}{0}
      \gridLine{-0.55}{1}{0.55}{1}
      \mpsC{0}{0}{1}{colMPS}
      \node at (-0.875,1) {\scalebox{0.8}{$s_1$}};
      \node at (0.875,1) {\scalebox{0.8}{$b_1$}};
      \node at (-0.875,0) {\scalebox{0.8}{$s_2$}};
      \node at (0.875,0) {\scalebox{0.8}{$b_2$}};
    \end{tikzpicture},
  \end{eqnarray}
  while for $s_1=s_2=1$ one needs to multiply the relevant tensors with a transformation analogous to~\eref{eq:defPentagon}
  \begin{eqnarray}
    \begin{tikzpicture}[baseline={([yshift=-0.6ex]current bounding box.center)},scale=0.5]
      \mpsWire{0}{-0.625}{0}{1.625}
      \gridLine{-0.55}{0}{0.55}{0}
      \gridLine{-0.55}{1}{0.55}{1}
      \mpsC{0}{0}{1}{colMPSBeta}
      \node at (-0.875,1) {\scalebox{0.8}{$1$}};
      \node at (0.875,1) {\scalebox{0.8}{$1$}};
      \node at (-0.875,0) {\scalebox{0.8}{$s$}};
      \node at (0.875,0) {\scalebox{0.8}{$b$}};
    \end{tikzpicture}=
    \begin{tikzpicture}[baseline={([yshift=-1.2ex]current bounding box.center)},scale=0.5]
      \mpsWire{0}{-0.625}{0}{2}
      \gridLine{-0.55}{0}{0.55}{0}
      \gridLine{-0.55}{1}{0.55}{1}
      \mpsC{0}{0}{1}{colMPS}
      \node at (-0.875,1) {\scalebox{0.8}{$1$}};
      \node at (0.875,1) {\scalebox{0.8}{$1$}};
      \node at (-0.875,0) {\scalebox{0.8}{$s$}};
      \node at (0.875,0) {\scalebox{0.8}{$b$}};
      \mpsAuxI{0}{1.625}{colMPSBeta}
    \end{tikzpicture},\qquad
    \begin{tikzpicture}[baseline={([yshift=-0.6ex]current bounding box.center)},scale=0.5]
      \mpsWire{0}{-0.5}{0}{0.5}
      \mpsAuxI{0}{0}{colMPSBeta}
    \end{tikzpicture}=
    \left[\matrix{
      \gamma & 0 & 0 \cr
      0 & 1 & 0 \cr
      0 & 0 & 1
      }\right],\qquad
      \begin{tikzpicture}[baseline={([yshift=-0.6ex]current bounding box.center)},scale=0.5]
        \mpsWire{0}{-0.5}{0}{1.25}
        \mpsAux{0}{0}{colMPSBeta}
        \mpsAuxI{0}{0.75}{colMPSBeta}
      \end{tikzpicture}=
      \begin{tikzpicture}[baseline={([yshift=-0.6ex]current bounding box.center)},scale=0.5]
        \mpsWire{0}{-0.5}{0}{1.25}
        \mpsAuxI{0}{0}{colMPSBeta}
        \mpsAux{0}{0.75}{colMPSBeta}
      \end{tikzpicture}=
      \gamma
      \begin{tikzpicture}[baseline={([yshift=-0.6ex]current bounding box.center)},scale=0.5]
        \mpsWire{0}{-0.5}{0}{1.25}
      \end{tikzpicture}.
  \end{eqnarray}
  Note that this holds for any label $\wLt$, $\wRt$, and $\wPlust$ upon replacing $\gamma$ with $e^{-i\beta}$, $e^{i\beta}$, and $e^{i\beta} \Lambda_{\beta}^{(+)}$ respectively.

  \subsection{Boundary tensors}\label{sec:tensorBoundary}
  The bottom boundary tensors are all the same and do not depend on the value of $\beta$,
  \begin{eqnarray}
    \begin{tikzpicture}[baseline={([yshift=-0.6ex]current bounding box.center)},scale=0.5]
      \mpsWire{0}{0}{0}{0.625}
      \mpsBvecV{0}{0}{colMPSBeta}
      \wedgeL{0}{0}
    \end{tikzpicture}=
    \begin{tikzpicture}[baseline={([yshift=-0.6ex]current bounding box.center)},scale=0.5]
      \mpsWire{0}{0}{0}{0.625}
      \mpsBvecV{0}{0}{colMPSBeta}
      \wedgeR{0}{0}
    \end{tikzpicture}=
    \begin{tikzpicture}[baseline={([yshift=-0.6ex]current bounding box.center)},scale=0.5]
      \mpsWire{0}{0}{0}{0.625}
      \mpsBvecV{0}{0}{colMPSBeta}
      \plus{0}{0}
    \end{tikzpicture}=
    \begin{tikzpicture}[baseline={([yshift=-0.6ex]current bounding box.center)},scale=0.5]
      \mpsWire{0}{0}{0}{0.625}
      \mpsBvecV{0}{0}{colMPS}
    \end{tikzpicture}=
    \left[\matrix{
      1-\vartheta \cr \vartheta \cr \frac{-\vartheta^2}{1-\vartheta}
    }\right]\mkern-6mu,\qquad
    \begin{tikzpicture}[baseline={([yshift=-0.6ex]current bounding box.center)},scale=0.5]
      \mpsWire{0}{0}{0}{0.625}
      \mpsBvecW{0}{0}{colMPSBeta}
      \wedgeL{0}{0}
    \end{tikzpicture}=
    \begin{tikzpicture}[baseline={([yshift=-0.6ex]current bounding box.center)},scale=0.5]
      \mpsWire{0}{0}{0}{0.625}
      \mpsBvecW{0}{0}{colMPSBeta}
      \wedgeR{0}{0}
    \end{tikzpicture}=
    \begin{tikzpicture}[baseline={([yshift=-0.6ex]current bounding box.center)},scale=0.5]
      \mpsWire{0}{0}{0}{0.625}
      \mpsBvecW{0}{0}{colMPSBeta}
      \plus{0}{0}
    \end{tikzpicture}=
    \begin{tikzpicture}[baseline={([yshift=-0.6ex]current bounding box.center)},scale=0.5]
      \mpsWire{0}{0}{0}{0.625}
      \mpsBvecW{0}{0}{colMPS}
    \end{tikzpicture}=
    \left[\matrix{
      1 \cr 0 \cr 0
    }\right]\mkern-6mu.
  \end{eqnarray}
  On the other hand, the top boundary vectors act on both the auxiliary space of
  the MPO describing $e^{i\beta Q^{(\pm)}}$, and the auxiliary space of the fixed-point MPO,
  and therefore must obtain the dependence on the charge in question and the value of $\beta$.
  In particular, the top boundary vector in the absence of the charge is
  \begin{eqnarray}\label{eq:topNoCharge}
    \begin{tikzpicture}[baseline={([yshift=-0.6ex]current bounding box.center)},scale=0.5]
      \node at (-0.75,0) {\scalebox{0.8}{$\phantom{s}$}};
      \mpsWire{0}{0}{0}{-0.625}
      \mpsBvec{0}{0}
    \end{tikzpicture}=\left[\matrix{1&1&0}\right],
  \end{eqnarray}
  while the top boundary vector included in $\sbra{l_{\beta,t}^{(-)}}$ is
  \begin{eqnarray}
    \begin{tikzpicture}[baseline={([yshift=-0.6ex]current bounding box.center)},scale=0.5]
      \mpsWire{0}{0}{0}{-0.625}
      \vLine{0}{0}{0.625}{0}
      \mpsBvecV{0}{0}{colMPSBeta}
      \wedgeL{0}{0}
      \node at (0.875,0) {\scalebox{0.8}{$x$}};
    \end{tikzpicture}\mkern-8mu=
    \begin{tikzpicture}[baseline={([yshift=-0.6ex]current bounding box.center)},scale=0.5]
      \mpsWire{0}{0}{0}{-0.625}
      \vLine{0}{0}{0.625}{0}
      \mpsBvecW{0}{0}{colMPSBeta}
      \wedgeL{0}{0}
      \node at (0.875,0) {\scalebox{0.8}{$x$}};
    \end{tikzpicture}\mkern-8mu=
    \left[\matrix{
      \delta_{x,1} & \delta_{x,2} & 0
    }\right].
  \end{eqnarray}
  Unlike the bulk tensors (cf.\ \eref{eq:newFixedPointTensors}) the top boundary
  vectors of $\sket{r_{\beta,t}^{(-)}}$ are not obtained just by the flip of the
  sign $\beta\mapsto -\beta$, but take a completely different form,
  \begin{eqnarray}
    \begin{tikzpicture}[baseline={([yshift=-0.6ex]current bounding box.center)},scale=0.5]
      \mpsWire{0}{0}{0}{-0.625}
      \vLine{0}{0}{-0.625}{0}
      \mpsBvecV{0}{0}{colMPSBeta}
      \wedgeR{0}{0}
      \node at (-0.875,0) {\scalebox{0.8}{$x$}};
    \end{tikzpicture}=
    \left[\matrix{
      1 & \delta_{x,1} + e^{i\beta} \delta_{x,2}& 0
    }\right],\qquad
    \begin{tikzpicture}[baseline={([yshift=-0.6ex]current bounding box.center)},scale=0.5]
      \mpsWire{0}{0}{0}{-0.625}
      \vLine{0}{0}{-0.625}{0}
      \mpsBvecW{0}{0}{colMPSBeta}
      \wedgeR{0}{0}
      \node at (-0.875,0) {\scalebox{0.8}{$x$}};
    \end{tikzpicture}=
    \left[\matrix{
      1 & \delta_{x,1} + e^{-i\beta} \delta_{x,2} & 0
    }\right].
  \end{eqnarray}
  We note that the overlap between the top boundary vectors
  of fixed-points and the boundary vectors from~\eref{eq:boundaryMPO} reduce to the
  top boundary vector of the fixed-point $\sbra{l_t}$,
  \begin{eqnarray}
    \begin{tikzpicture}[baseline={([yshift=-0.6ex]current bounding box.center)},scale=0.5]
      \mpsWire{0}{0}{0}{-0.625}
      \vLine{0}{0}{-0.625}{0}
      \vLine{-0.625}{-0.25}{-0.625}{0.25}
      \mpsBvecV{0}{0}{colMPSBeta}
      \wedgeR{0}{0}
    \end{tikzpicture}=
    \begin{tikzpicture}[baseline={([yshift=-0.6ex]current bounding box.center)},scale=0.5]
      \mpsWire{0}{0}{0}{-0.625}
      \vLine{0}{0}{-0.625}{0}
      \vLine{-0.625}{-0.25}{-0.625}{0.25}
      \mpsBvecW{0}{0}{colMPSBeta}
      \wedgeR{0}{0}
    \end{tikzpicture}=
    \begin{tikzpicture}[baseline={([yshift=-0.6ex]current bounding box.center)},scale=0.5]
      \mpsWire{0}{0}{0}{-0.625}
      \vLine{0}{0}{0.625}{0}
      \vLine{0.625}{-0.25}{0.625}{0.25}
      \mpsBvecV{0}{0}{colMPSBeta}
      \wedgeL{0}{0}
    \end{tikzpicture}=
    \begin{tikzpicture}[baseline={([yshift=-0.6ex]current bounding box.center)},scale=0.5]
      \mpsWire{0}{0}{0}{-0.625}
      \vLine{0}{0}{0.625}{0}
      \vLine{0.625}{-0.25}{0.625}{0.25}
      \mpsBvecW{0}{0}{colMPSBeta}
      \wedgeL{0}{0}
    \end{tikzpicture}=
    \begin{tikzpicture}[baseline={([yshift=-0.6ex]current bounding box.center)},scale=0.5]
      \mpsWire{0}{0}{0}{-0.625}
      \mpsBvec{0}{0}
    \end{tikzpicture}.
  \end{eqnarray}

  In the case of $Q^{(+)}$, the top boundary vectors take the following form,
  \begin{eqnarray}\fl
    \eqalign{
      &\mkern-11mu
      \begin{tikzpicture}[baseline={([yshift=-0.6ex]current bounding box.center)},scale=0.5]
        \mpsWire{0}{0}{0}{-0.625}
        \vLine{0}{0}{-0.625}{0}
        \mpsBvecW{0}{0}{colMPSBeta}
        \plus{0}{0}
        \node at (-0.875,0) {\scalebox{0.8}{$x$}};
      \end{tikzpicture}=
      \left[\matrix{
        \delta_{x,1}+e^{2i\beta}\delta_{x,2}+\delta_{x,3}\cr
        \delta_{x,1}+\frac{e^{2i\beta}}{\gamma_{+}}(\delta_{x,2}+\delta_{x,3})\cr
        \left(\frac{e^{3i\beta}}{\gamma_{+}}-1\right)\delta_{x,1}
      }\right]^{T}\mkern-10mu,\\
      &\begin{tikzpicture}[baseline={([yshift=-0.6ex]current bounding box.center)},scale=0.5]
        \mpsWire{0}{0}{0}{-0.625}
        \vLine{0}{0}{0.625}{0}
        \mpsBvecW{0}{0}{colMPSBeta}
        \plus{0}{0}
        \node at (0.875,0) {\scalebox{0.8}{$x$}};
      \end{tikzpicture}\mkern-8mu=
      \left[\matrix{
        \frac{e^{i\beta}}{\gamma_{+}} \delta_{x,1}\cr 
        \frac{e^{2i\beta}(1-\vartheta)}{\gamma_{+}^2} \left(\frac{\vartheta}{1-\vartheta}\delta_{x,2}+\delta_{x,3}\right)\cr 
        \frac{e^{2i\beta}(1-\vartheta)}{\gamma_{+}^2}\left(\delta_{x,2}-\delta_{x,3}\right)
      }\right]^{T}\mkern-10mu,}
      \qquad
      \eqalign{
        &\mkern-4mu
        \begin{tikzpicture}[baseline={([yshift=-0.6ex]current bounding box.center)},scale=0.5]
          \mpsWire{0}{0}{0}{-0.625}
          \vLine{0}{0}{-0.625}{0}
          \mpsBvecV{0}{0}{colMPSBeta}
          \plus{0}{0}
          \node at (-0.875,0) {\scalebox{0.8}{$x$}};
        \end{tikzpicture}=
        \left[\matrix{
          \delta_{x,1}+e^{2i\beta}\delta_{x,2}+\delta_{x,3}\cr
          e^{2i\beta}\delta_{x,1}+\gamma_{+}(\delta_{x,2}+\delta_{x,3})\cr
          \left(e^{3i\beta}-\gamma_{+}\right)(\delta_{x,2}+\delta_{x,3})
        }\right]^T\mkern-10mu,\\ 
        &\begin{tikzpicture}[baseline={([yshift=-0.6ex]current bounding box.center)},scale=0.5]
          \mpsWire{0}{0}{0}{-0.625}
          \vLine{0}{0}{0.625}{0}
          \mpsBvecV{0}{0}{colMPSBeta}
          \plus{0}{0}
          \node at (0.875,0) {\scalebox{0.8}{$x$}};
        \end{tikzpicture}\mkern-8mu=
        \left[\matrix{
          \frac{e^{i\beta}}{\gamma_{+}} \delta_{x,1}\cr 
          \frac{e^{i\beta}(1-\vartheta)}{\gamma_{+}} 
          \left(\delta_{x,2}+\frac{\vartheta e^{2i\beta}}{1-\vartheta}\delta_{x,3}\right)\cr 
          \frac{e^{i\beta}(1-\vartheta)}{\gamma_{+}} 
          \left(-\delta_{x,2}+e^{2i\beta}\delta_{x,3}\right)
        }\right]^T\mkern-10mu,}
  \end{eqnarray}
  where 
  \begin{eqnarray}
    \gamma_{+}=e^{i\beta}\Lambda_{\beta}^{(+)},
  \end{eqnarray}
  and $\Lambda_{\beta}^{(+)}$ is given in Eq.~\eref{eq:defLambda}. The two overlaps
  between these vectors and the boundary vectors of the $Q^{(+)}$ MPO that are needed
  to evaluate Eq.~\eref{eq:defTNOneEdgeContrib} are given as
  \begin{eqnarray}
    \begin{tikzpicture}[baseline={([yshift=-0.6ex]current bounding box.center)},scale=0.5]
      \mpsWire{0}{0}{0}{-0.625}
      \vLine{0}{0}{0.625}{0}
      \vLine{0.625}{-0.25}{0.625}{0.25}
      \mpsBvecW{0}{0}{colMPSBeta}
      \plus{0}{0}
    \end{tikzpicture}=
    \left[\matrix{
      \frac{e^{i\beta}}{\gamma_{+}}\cr
      \frac{e^{i\beta}(1-\vartheta+\vartheta e^{i\beta})}{\gamma_{+}^2}\cr
      -\frac{e^{i\beta}(1-e^{i\beta})(1-\vartheta)}{\gamma_{+}^2}
    }\right]^{T},\qquad
    \begin{tikzpicture}[baseline={([yshift=-0.6ex]current bounding box.center)},scale=0.5]
      \mpsWire{0}{0}{0}{-0.625}
      \vLine{0}{0}{-0.625}{0}
      \vLine{-0.625}{-0.25}{-0.625}{0.25}
      \mpsBvecV{0}{0}{colMPSBeta}
      \plus{0}{0}
    \end{tikzpicture}=
    \left[\matrix{
      1\cr
      \gamma_{+} e^{-i\beta}\cr
      e^{2 i\beta}-\gamma_{+} e^{-i\beta}
    }\right]^{T}.
  \end{eqnarray}
  Note that in the limit of $\beta\to 0$ all of the overlaps reduce to
  \begin{eqnarray}
    \left.
    \begin{tikzpicture}[baseline={([yshift=-0.6ex]current bounding box.center)},scale=0.5]
      \mpsWire{0}{0}{0}{-0.625}
      \vLine{0}{0}{0.625}{0}
      \vLine{0.625}{-0.25}{0.625}{0.25}
      \mpsBvecV{0}{0}{colMPSBeta}
      \plus{0}{0}
    \end{tikzpicture}\right|_{\beta\to 0}
    =
    \left.
    \begin{tikzpicture}[baseline={([yshift=-0.6ex]current bounding box.center)},scale=0.5]
      \mpsWire{0}{0}{0}{-0.625}
      \vLine{0}{0}{0.625}{0}
      \vLine{0.625}{-0.25}{0.625}{0.25}
      \mpsBvecW{0}{0}{colMPSBeta}
      \plus{0}{0}
    \end{tikzpicture}\right|_{\beta\to 0}
    =
    \left.
    \begin{tikzpicture}[baseline={([yshift=-0.6ex]current bounding box.center)},scale=0.5]
      \mpsWire{0}{0}{0}{-0.625}
      \vLine{0}{0}{-0.625}{0}
      \vLine{-0.625}{-0.25}{-0.625}{0.25}
      \mpsBvecV{0}{0}{colMPSBeta}
      \plus{0}{0}
    \end{tikzpicture}\right|_{\beta\to0}
    =
    \begin{tikzpicture}[baseline={([yshift=-0.2ex]current bounding box.center)},scale=0.5]
      \mpsWire{0}{0}{0}{-1}
      \vLine{0}{0}{-0.625}{0}
      \vLine{-0.625}{-0.25}{-0.625}{0.25}
      \mpsBvecW{0}{0}{colMPSBeta}
      \plus{0}{0}
      \mpsAux{0}{-0.625}{colMPSBeta}
      \plus{0}{-0.625}
    \end{tikzpicture}
    \mkern-4mu
    \left.
    \vphantom{
    \begin{tikzpicture}[baseline={([yshift=-0.6ex]current bounding box.center)},scale=0.5]
      \mpsWire{0}{0}{0}{-0.625}
      \vLine{0}{0}{0.625}{0}
      \vLine{0.625}{-0.25}{0.625}{0.25}
      \mpsBvecW{0}{0}{colMPSBeta}
      \plus{0}{0}
    \end{tikzpicture}}\right|_{\beta\to 0}=
    \begin{tikzpicture}[baseline={([yshift=-0.6ex]current bounding box.center)},scale=0.5]
      \mpsWire{0}{0}{0}{-0.625}
      \mpsBvec{0}{0}
    \end{tikzpicture}.
  \end{eqnarray}
  \subsection{Normalization}\label{sec:FPnormalization}
  The normalization of boundary vectors above was chosen so that the following holds,
  \begin{eqnarray}
    \begin{tikzpicture}[baseline={([yshift=-0.6ex]current bounding box.center)},scale=0.5]
      \mpsWire{0}{2}{0}{3}
      \mpsWire{1}{2}{1}{3}
      \vLine{0}{3}{1}{3}
      \mpsBvecW{0}{3}{colMPSBeta}
      \mpsBvecV{1}{3}{colMPSBeta}
      \mpsBvecW{0}{2}{colMPS}
      \mpsBvecV{1}{2}{colMPS}
      \foreach \x in {3}{
        \plus{0}{\x}
        \plus{1}{\x}
      }
    \end{tikzpicture}=
    \begin{tikzpicture}[baseline={([yshift=-0.6ex]current bounding box.center)},scale=0.5]
      \mpsWire{0}{2}{0}{3}
      \mpsWire{1}{2}{1}{3}
      \vLine{0}{3}{1}{3}
      \mpsBvecV{0}{3}{colMPSBeta}
      \mpsBvecW{1}{3}{colMPSBeta}
      \mpsBvecV{0}{2}{colMPS}
      \mpsBvecW{1}{2}{colMPS}
      \foreach \x in {3}{
        \plus{0}{\x}
        \plus{1}{\x}
      }
    \end{tikzpicture}=
    \begin{tikzpicture}[baseline={([yshift=-0.6ex]current bounding box.center)},scale=0.5]
      \mpsWire{0}{2}{0}{3}
      \mpsWire{1}{2}{1}{3}
      \vLine{0}{3}{1}{3}
      \mpsBvecW{0}{3}{colMPSBeta}
      \mpsBvecV{1}{3}{colMPSBeta}
      \mpsBvecW{0}{2}{colMPS}
      \mpsBvecV{1}{2}{colMPS}
      \foreach \x in {3}{
        \wedgeL{0}{\x}
        \wedgeR{1}{\x}
      }
    \end{tikzpicture}=
    \begin{tikzpicture}[baseline={([yshift=-0.6ex]current bounding box.center)},scale=0.5]
      \mpsWire{0}{2}{0}{3}
      \mpsWire{1}{2}{1}{3}
      \vLine{0}{3}{1}{3}
      \mpsBvecV{0}{3}{colMPSBeta}
      \mpsBvecW{1}{3}{colMPSBeta}
      \mpsBvecV{0}{2}{colMPS}
      \mpsBvecW{1}{2}{colMPS}
      \foreach \x in {3}{
        \wedgeL{0}{\x}
        \wedgeR{1}{\x}
      }
    \end{tikzpicture}=1,
  \end{eqnarray}
  which together with 
  \begin{eqnarray}\fl
    \begin{tikzpicture}[baseline={([yshift=-0.6ex]current bounding box.center)},scale=0.5]
      \draw[thick,colLines,rounded corners] (0,2) -- (-0.75,2) -- (-0.75,3.5) -- (1.75,3.5) -- (1.75,2) -- cycle;
      \mpsWire{0}{1.5}{0}{3}
      \mpsWire{1}{1.5}{1}{3}
      \vLine{0}{3}{1}{3}
      \mpsBvecW{0}{3}{colMPSBeta}
      \mpsB{0}{2}{colMPSBeta}
      \mpsA{1}{2}{colMPSBeta}
      \mpsBvecV{1}{3}{colMPSBeta}
      \foreach \x in {2,3}{
        \plus{0}{\x}
        \plus{1}{\x}
      }
    \end{tikzpicture}=\Lambda_{\beta}^{(+)}
    \begin{tikzpicture}[baseline={([yshift=-0.6ex]current bounding box.center)},scale=0.5]
      \mpsWire{0}{2.25}{0}{3}
      \mpsWire{1}{2.25}{1}{3}
      \vLine{0}{3}{1}{3}
      \mpsBvecV{0}{3}{colMPSBeta}
      \mpsBvecW{1}{3}{colMPSBeta}
      \foreach \x in {3}{
        \plus{0}{\x}
        \plus{1}{\x}
      }
    \end{tikzpicture},\quad
    \begin{tikzpicture}[baseline={([yshift=-0.6ex]current bounding box.center)},scale=0.5]
      \draw[thick,colLines,rounded corners] (0,2) -- (-0.75,2) -- (-0.75,3.5) -- (1.75,3.5) -- (1.75,2) -- cycle;
      \mpsWire{0}{1.5}{0}{3}
      \mpsWire{1}{1.5}{1}{3}
      \vLine{0}{3}{1}{3}
      \mpsBvecV{0}{3}{colMPSBeta}
      \mpsA{0}{2}{colMPSBeta}
      \mpsB{1}{2}{colMPSBeta}
      \mpsBvecW{1}{3}{colMPSBeta}
      \foreach \x in {2,3}{
        \plus{0}{\x}
        \plus{1}{\x}
      }
    \end{tikzpicture}=\Lambda_{\beta}^{(+)}
    \begin{tikzpicture}[baseline={([yshift=-0.6ex]current bounding box.center)},scale=0.5]
      \mpsWire{0}{2.25}{0}{3}
      \mpsWire{1}{2.25}{1}{3}
      \vLine{0}{3}{1}{3}
      \mpsBvecW{0}{3}{colMPSBeta}
      \mpsBvecV{1}{3}{colMPSBeta}
      \foreach \x in {3}{
        \plus{0}{\x}
        \plus{1}{\x}
      }
    \end{tikzpicture},\quad
    \begin{tikzpicture}[baseline={([yshift=-0.6ex]current bounding box.center)},scale=0.5]
      \draw[thick,colLines,rounded corners] (0,2) -- (-0.75,2) -- (-0.75,3.5) -- (1.75,3.5) -- (1.75,2) -- cycle;
      \mpsWire{0}{1.5}{0}{3}
      \mpsWire{1}{1.5}{1}{3}
      \vLine{0}{3}{1}{3}
      \mpsBvecW{0}{3}{colMPSBeta}
      \mpsB{0}{2}{colMPSBeta}
      \mpsA{1}{2}{colMPSBeta}
      \mpsBvecV{1}{3}{colMPSBeta}
      \foreach \x in {2,3}{
        \wedgeL{0}{\x}
        \wedgeR{1}{\x}
      }
    \end{tikzpicture}=
    \begin{tikzpicture}[baseline={([yshift=-0.6ex]current bounding box.center)},scale=0.5]
      \mpsWire{0}{2.25}{0}{3}
      \mpsWire{1}{2.25}{1}{3}
      \vLine{0}{3}{1}{3}
      \mpsBvecV{0}{3}{colMPSBeta}
      \mpsBvecW{1}{3}{colMPSBeta}
      \foreach \x in {3}{
        \wedgeL{0}{\x}
        \wedgeR{1}{\x}
      }
    \end{tikzpicture},\quad
    \begin{tikzpicture}[baseline={([yshift=-0.6ex]current bounding box.center)},scale=0.5]
      \draw[thick,colLines,rounded corners] (0,2) -- (-0.75,2) -- (-0.75,3.5) -- (1.75,3.5) -- (1.75,2) -- cycle;
      \mpsWire{0}{1.5}{0}{3}
      \mpsWire{1}{1.5}{1}{3}
      \vLine{0}{3}{1}{3}
      \mpsBvecV{0}{3}{colMPSBeta}
      \mpsA{0}{2}{colMPSBeta}
      \mpsB{1}{2}{colMPSBeta}
      \mpsBvecW{1}{3}{colMPSBeta}
      \foreach \x in {2,3}{
        \wedgeL{0}{\x}
        \wedgeR{1}{\x}
      }
    \end{tikzpicture}=
    \begin{tikzpicture}[baseline={([yshift=-0.6ex]current bounding box.center)},scale=0.5]
      \mpsWire{0}{2.25}{0}{3}
      \mpsWire{1}{2.25}{1}{3}
      \vLine{0}{3}{1}{3}
      \mpsBvecW{0}{3}{colMPSBeta}
      \mpsBvecV{1}{3}{colMPSBeta}
      \foreach \x in {3}{
        \wedgeL{0}{\x}
        \wedgeR{1}{\x}
      }
    \end{tikzpicture},
  \end{eqnarray}
  implies
  \begin{eqnarray}
    \sbraket{l_{\beta,t}^{(+)}}{r_{\beta,t}^{(+)}}=\left.\Lambda_{\beta}^{(+)}\right.^{2t},
    \qquad
    \sbraket{l_{\beta,t}^{(-)}}{r_{\beta,t}^{(-)}}=1.
  \end{eqnarray}

\end{document}